\documentclass[11pt]{article}
\def\one{1\hskip -.37em 1}     

\begin{document}
\begin{titlepage}
\begin{flushright}
hep-ph/0201280\\
January 2002
\end{flushright}
\begin{centering}
 
{\ }\vspace{1cm}
 
{\Large\bf Neutrino-Charged Matter Interactions:}\\

\vspace{10pt}

{\Large\bf a General Four-Fermion Effective Parametrization}\\

\vspace{2.0cm}

Jean El Bachir Mendy

\vspace{0.3cm}

{\em Institut de Math\'ematiques et des Sciences Physiques (IMSP)}\\
{\em Unit\'e de Recherche en Physique Th\'eorique,
Universit\'e Nationale du B\'enin}\\
{\em B.P. 2628 , Porto-Novo, Republic of Benin}\\
{\tt mendiz19@yahoo.fr}

\vspace{0.6cm}

Jan Govaerts

\vspace{0.3cm}

{\em Institute of Nuclear Physics}\\
{\em Catholic University of Louvain}\\
{\em 2, Chemin du Cyclotron, B-1348 Louvain-la-Neuve, Belgium}\\
{\tt govaerts@fynu.ucl.ac.be}

\vspace{1.5cm}
\begin{abstract}

\noindent Given the eventuality of neutrino and muon factories in the 
foreseeable future, all possible $2\rightarrow 2$ processes involving two 
neutrinos, whether Dirac or Majorana ones, and two charged fermions are 
considered on the basis of the most general Lorentz invariant four-fermion 
effective interaction possible, in the limit of massless particles. 
Such a parametrization should enable the assessment of the sensitivity 
to physics beyond the Standard Model, including the eventual discrimination 
between the Dirac or Majorana character of neutrinos,
of specific experimental beam and detector designs.

\end{abstract}

\vspace{10pt}

\end{centering} 

\vspace{5pt}


\vspace{25pt}

\end{titlepage}

\setcounter{footnote}{0}

\section{\bf Introduction.}
\label{Sect1}

As is widely appreciated, the most general Lorentz invariant four-fermion 
effective parametrization of electroweak processes has played a central
role in unravelling the basic structure and chiral properties of this
fundamental interaction. Still to this day, the original analysis of
Ref.\cite{JTW} is used in precision studies of $\beta$-decay\cite{DQ,RPP},
aiming at identifying at low energies tell-tale signs for physics beyond 
the Standard Model (SM). Likewise at intermediate and high energies in 
the purely muon and tau leptonic sectors, a similar parametrization\cite{WF}
has become the standard\cite{RPP} in terms of which to confine ever further 
parameter space, hoping to uncover a lack of overlap with that of the SM. 
A similar approach is also possible for precision studies of semi-leptonic 
processes which involve both the muon sector and the first $(u,d)$ quark 
generation, for example at the intermediate energies of nuclear muon 
capture\cite{JG}.

With the foreseen advent of neutrino factories and muon colliders, an
analogous general analysis, involving in particular neutrino beams, appears
to be of potential interest in the design of eventual experiments and 
detectors. For instance, the possibility of intersecting neutrino beams 
should not be dismissed, especially in the eventuality of very large 
intensities. Indeed, in spite of small rates, if only a single 
$\nu_a\nu_b\rightarrow\ell^-_i\ell^+_j$ event for instance---as opposed to
$\nu_a\bar{\nu}_b\rightarrow\ell^-_i\ell^+_j$---were to be observed,
lepton number violation would definitely have been
established, which most likely would imply the Majorana character
of neutrinos, one of the most pressing issues in neutrino physics today.

This note presents such an analysis, based on the most general four-fermion
effective interaction possible of two neutrinos and two charged fermions 
(whether leptons or quarks) of fixed ``flavours", or rather more correctly,
of definite mass eigenstates, solely constrained by the requirements of
Lorentz invariance and electric charge conservation. For instance, even though
this might be realized only in small and peculiar classes of models beyond 
the SM, allowance is made for the possibility that both the neutrino fields 
and their charge conjugates couple in the effective Lagrangian density. 
Furthermore, the analysis is developed separately whether for Dirac or 
Majorana neutrinos, with the hope to identify circumstances under which 
scattering experiments involving neutrinos could help discriminate between 
these two cases through different angular correlations for differential 
cross sections, given the high rates to be expected at neutrino
factories. As is well known, 
the ``practical Dirac-Majorana confusion theorem"\cite{Confusion} states
that within the SM, namely in the limit of massless neutrinos and
$(V-A)$ interactions only, these two possibilities are physically totally
equivalent, and hence cannot be distinguished. On the other hand,
relaxing the purely $(V-A)$ structure of the electroweak interaction by 
including at least another interaction whose chirality structure is different
should suffice to evade this conclusion, even in the limit of massless 
neutrinos. 

The general classes of processes considered in this note comprise
neutrino pair annihilation into charged leptons\footnote{Henceforth, the
charged fermions are referred to as leptons, even though exactly the same
analysis and results apply to quark states, with due account then for the
quark colour degree of freedom and the quark structure of the hadrons involved. 
Also, charged leptons will simply be called leptons, for short.}, the inverse 
process of neutrino pair production through lepton annihilation, and finally 
neutrino-lepton scattering. These processes will also be considered whether 
either one or both pairs of neutrino and lepton flavours\footnote{In fact, 
our analysis considers specific mass eigenstates for the external neutrino 
and lepton states, in the massless limit, namely when all other energy scales 
are much larger than the masses of these particles. By abuse of language, we 
shall refer to these mass eigenstates as ``flavour" ones, following a widespread
usage.}, $(a,b)$ and $(i,j)$ respectively, are identical or not. The 
sole implicit assumption is that the energy available to the reaction is both
sufficiently large in order to justify ignoring neutrino and lepton masses,
and sufficiently small in order to justify the four-fermion parametrization
of the boson exchanges responsible for the interactions. 
In other words, the calculations are all performed in the limit of zero mass
for all external neutrino and lepton mass eigenstates.
Nonetheless, effects that distinguish Majorana from Dirac neutrinos 
should survive in this massless limit. In addition to being much larger than 
the neutrino and lepton mass scales, the energy scale available to
the process must also be much smaller than the energy scales associated to 
the interactions modeled by the four-fermion interactions. 

In terms of classes of processes, our analysis thus covers a wide variety 
of possibilities, and its results are presented in a manner which, it is hoped, 
will be found readily useful for implementation in numerical codes, whatever
a specific model for physics beyond the SM, namely a specific set of
effective couplings parametrizing the four-fermion interactions.
In this note, no attempt is made at developing a systematic analysis
to assess the physics potential of specific processes based on some
particular beam and detector design, whether to look for physics beyond the
SM, or discriminate between the Dirac or Majorana character of neutrinos. 
The main purpose of this work is to provide the general parametrization that 
is required for such a dedicated assessment, left for future analysis.

The note is organized as follows. The next section provides what might be 
called the ``kinematics" of the analysis, by recalling some simple facts about
Dirac and Majorana fermions. Sect.\ref{Sect3} discusses the general
four-fermion effective Lagrangian used in our analysis. Sects.\ref{Sect4} to
\ref{Sect6} then list the results for the three classes of processes
mentioned above, with Sect.\ref{Sect7} only superficially illustrating the
potential reach of these types of processes at neutrino factories.
Concluding remarks are presented in Sect.\ref{Sect8}.

\section{A Compendium of Simple Properties}
\label{Sect2}

\subsection{Dirac, Weyl and Majorana spinors}
\label{Subsect2.1}

The purpose of this section is to recall a series of results relevant
to Dirac, Weyl and Majorana quantum fermionic fields, and to specify our
conventions. Since all processes are considered in the limit of massless 
neutrinos and leptons, the representation of the Clifford-Dirac algebra 
$\{\gamma^\mu,\gamma^\nu\}=2g^{\mu\nu}$ used througout is the
chiral one, which we take to be
\begin{equation}
\gamma^0=\left(\begin{array}{c c}
0 & -\one \\
-\one & 0 \end{array}\right)\ \ ,\ \ 
\gamma^i=\left(\begin{array}{c c}
0 & \sigma^i \\
-\sigma^i & 0 \end{array}\right)\ ,\ i=1,2,3,\ \ ,\ \ 
\gamma_5=i\gamma^0\gamma^1\gamma^2\gamma^3=\left(\begin{array}{c c}
\one & 0 \\
0 & -\one \end{array}\right)\ \ ,
\end{equation}
$\sigma^i$ ($i=1,2,3$) being of course the usual Pauli matrices
(our choice of Minkowski metric signature $g_{\mu\nu}$ is $(+---)$). 
The chiral projectors $P_\eta$ ($\eta=\pm$) are given by
\begin{equation}
P_\eta=\frac{1}{2}\left[1+\eta\gamma_5\right]\ \ ,\ \ 
P^2_\eta=P_\eta\ \ \ ,\ \ \ P_\eta P_{-\eta}=0\ \ ,\ \ \eta=+,-\ .
\end{equation}

By definition, the charge conjugation matrix $C$ is such that
\begin{equation}
C^{-1}\one C=\one^{\rm T}\ ,\
C^{-1}\gamma_5C=\gamma_5^{\rm T}\ ,\
C^{-1}\gamma^\mu C=-{\gamma^\mu}^{\rm T}\ ,\
C^{-1}\left(\gamma^\mu\gamma_5\right)C=
{\left(\gamma^\mu\gamma_5\right)}^{\rm T}\ ,\
\end{equation}
\begin{equation}
C^{-1}\sigma_{\mu\nu}C=-\sigma_{\mu\nu}^{\rm T}\ ,\
C^{-1}\left(\sigma_{\mu\nu}\gamma_5\right)C=
-\left(\sigma_{\mu\nu}\gamma_5\right)^{\rm T}\ ,\
\end{equation}
with
\begin{equation}C^{\rm T}=C^\dagger=-C\ \ \ ,\ \ \ 
CC^\dagger=\one=C^\dagger C\ ,
\end{equation}
and which, in the chiral representation, is given by
\begin{equation}
C=\left(\begin{array}{c c}
-i\sigma^2 & 0 \\
0 & i\sigma^2 \end{array}\right)\ .
\end{equation}

Given a four component Dirac spinor $\psi$, our definition of the associated
charge conjugate spinor is such that
\begin{equation}
\psi_c=\psi^c=\lambda\,C\overline{\psi}^{\rm T}\ ,
\end{equation}
where $\lambda$ is some arbitrary unit phase factor, whose value would
depend {\sl a priori\/} on the choice of spinor field ({\sl i.e.\/} on
the neutrino or lepton flavour hereafter). This freedom in the choice
of phase factor under charge conjugation is directly related to the
``creation phase factor" of Ref.\cite{Kayser}, as shown below.

Solutions to the free massless Dirac equation may be expanded in the helicity 
basis, in terms of the following mode representation of a Dirac quantum
spinor $\psi_D(x)$,
\begin{equation}
\psi_D(x)=\int_{(\infty)}\frac{d^3\vec{k}}{(2\pi)^32|\vec{k}|}\,
\sum_{\eta=\pm}\left[e^{-ik\cdot x}u(\vec{k},\eta)b(\vec{k},\eta)+
e^{ik\cdot x}v(\vec{k},\eta)d^\dagger(\vec{k},\eta)\right]\ .
\end{equation}
Here, the fermionic creation and annihilation operators have the Lorentz
covariant normalization
\begin{equation}
\left\{b(\vec{k},\eta),b^\dagger(\vec{k}',\eta')\right\}=
(2\pi)^32|\vec{k}|\delta_{\eta,\eta'}\delta^{(3)}(\vec{k}-\vec{k}')=
\left\{d(\vec{k},\eta),d^\dagger(\vec{k}',\eta')\right\}\ ,
\end{equation}
while the plane wave spinors $u(\vec{k},\eta)$ and $v(\vec{k},\eta)$ are
given by,
\begin{equation}
u(\vec{k},+)=v(\vec{k},-)=\sqrt{2|\vec{k}|}\left(\begin{array}{c}
\chi_+(\hat{k}) \\ 0 \end{array}\right)\ \ ,\ \ 
u(\vec{k},-)=v(\vec{k},+)=\sqrt{2|\vec{k}|}\left(\begin{array}{c}
0 \\ \chi_-(\hat{k}) \end{array}\right)\ \ ,
\end{equation}
with the Pauli bi-spinors
\begin{equation}
\chi_+(\hat{k})=\left(\begin{array}{c}
e^{-i\varphi/2}\cos\theta/2 \\ e^{i\varphi/2}\sin\theta/2 \end{array}\right)
\ \ ,\ \ 
\chi_-(\hat{k})=\left(\begin{array}{c}
-e^{-i\varphi/2}\sin\theta/2 \\ e^{i\varphi/2}\cos\theta/2 \end{array}\right)\ ,
\end{equation}
such that
\begin{equation}
\hat{k}\cdot\vec{\sigma}\,\chi_\eta(\hat{k})=\eta\,\chi_\eta(\hat{k})
\ \ \ ,\ \ \
\chi_\eta(\hat{k})\chi^\dagger_\eta(\hat{k})=\frac{1}{2}
\left[\one+\eta\hat{k}\cdot\vec{\sigma}\right]\ ,
\end{equation}
$\varphi$ and $\theta$ being of course the usual spherical angles for the
unit vector $\hat{k}=\vec{k}/|\vec{k}|$ with respect to the axes $i=1,2,3$,
namely $\hat{k}=(\sin\theta\cos\varphi,\sin\theta\sin\varphi,\cos\theta)$. 

The value of the index $\eta=\pm$ coincides with the helicity
of the corresponding massless one-particle states, and coincides of course 
with the chirality of the associated quantum field.
Namely, left- or right-handed four component Weyl spinors, with
$\eta=-$ et $\eta=+$ respectively, have the following mode decompositions
\begin{equation}
\psi_\eta(x)=\int_{(\infty)}\frac{d^3\vec{k}}{(2\pi)^32|\vec{k}|}\,
\left[e^{-ik\cdot x}u(\vec{k},\eta)b(\vec{k},\eta)+
e^{ik\cdot x}v(\vec{k},-\eta)d^\dagger(\vec{k},-\eta)\right]\ ,
\end{equation}
as implied by the identification
\begin{equation}
\psi_\eta(x)=P_\eta\,\psi_D(x)\ .
\end{equation}
Hence, $b^\dagger(\vec{k},\eta)$ and $d^\dagger(\vec{k},\eta)$ are the
creation operators of a particle and of an antiparticule, respectively, both
of helicity $\eta$ and momentum $\vec{k}$.

This identification may also be established from the chiral properties of
the plane wave spinors,
\begin{equation}
P_\eta\,u(\vec{k},\eta)=u(\vec{k},\eta)\ \ ,\ \ 
P_\eta\,u(\vec{k},-\eta)=0\ \ ;\ \ 
P_\eta\,v(\vec{k},\eta)=0\ \ ,\ \ 
P_\eta\,v(\vec{k},-\eta)=v(\vec{k},-\eta)\ \ , 
\end{equation}
\begin{equation}
\overline{u}(\vec{k},\eta)\,P_\eta=0\ \ ,\ \ 
\overline{u}(\vec{k},-\eta)\,P_\eta=\overline{u}(\vec{k},-\eta)\ \ ;\ \ 
\overline{v}(\vec{k},\eta)\,P_\eta=\overline{v}(\vec{k},\eta)\ \ ,\ \ 
\overline{v}(\vec{k},-\eta)\,P_\eta=0\ \ , 
\end{equation}
as well as
\begin{equation}
u(\vec{k},\eta)\overline{u}(\vec{k},\eta)=\frac{\one+\eta\gamma_5}{2}/\!\!\!k
\ \ \ ,\ \ \ 
v(\vec{k},\eta)\overline{v}(\vec{k},\eta)=\frac{\one-\eta\gamma_5}{2}/\!\!\!k
\ \ .
\end{equation}

Likewise, their properties under charge conjugation are such that
\begin{equation}
C\overline{u}^{\rm T}(\vec{k},\eta)=v(\vec{k},\eta)\ \ ,\ \ 
C\overline{v}^{\rm T}(\vec{k},\eta)=u(\vec{k},\eta)\ \ ;\ \ 
\overline{v}(\vec{k},\eta)=u^{\rm T}(\vec{k},\eta)C\ \ ,\ \ 
\overline{u}(\vec{k},\eta)=v^{\rm T}(\vec{k},\eta)C\ ,
\end{equation}
these results being specific to the helicity basis.
Charge conjugates of spinors are then given by, say for
a Dirac spinor $\psi_D(x)$,
\begin{equation}
\psi^c_D(x)=\int_{(\infty)}\frac{d^3\vec{k}}{(2\pi)^32|\vec{k}|}\,
\sum_{\eta=\pm}\left[e^{-ik\cdot x}\lambda u(\vec{k},\eta)d(\vec{k},\eta)+
e^{ik\cdot x}\lambda v(\vec{k},\eta)b^\dagger(\vec{k},\eta)\right]\ ,
\end{equation}
as one should expect of course.

Finally, let us turn to Majorana spinors. As opposed to a Dirac spinor
which is comprised of two independent Weyl spinors of opposite chiralities,
namely one of each of the two fundamental representations of the 
(covering group of the) Lorentz group,
\begin{equation}
\psi_D(x)=\psi_+(x)+\psi_-(x)\ ,
\end{equation}
a Majorana spinor $\psi_M(x)$ is a four component spinor which is covariant 
under Lorentz transformations but which is constructed this time from a single 
Weyl spinor, say of left-handed chirality\footnote{Since charge conjugation 
exchanges left- and right-handed chiralities, the chirality of the basic 
Weyl spinor used in this construction is irrelevant to the definition of a 
Majorana spinor.} $\eta=-$, and which is invariant under charge 
conjugation\footnote{Note that a similar definition starting from a Dirac 
rather than a Weyl spinor might be contemplated, then leading however to 
two independent Majorana spinors, each of which is obtained in the manner 
just described from a single distinct Weyl spinor, namely 
$\psi^{(1)}_M=(\psi_D+\psi^c_D)/\sqrt{2}$ and
$\psi^{(2)}_M=-i(\psi_D-\psi^c_D)/\sqrt{2}$, in complete analogy with the
real and imaginary parts of a single complex scalar field as well as the
physical interpretation of the associated quanta as being particles which
are or not their own antiparticles. Specifically, we have
$\psi^{(1)}_M=\psi^{(1)}_-+{\psi^{(1)}_-}^c$,
$\psi^{(2)}_M=\psi^{(2)}_-+{\psi^{(2)}_-}^c$ with
$\psi^{(1)}_-=(\psi_-+\psi^c_+)/\sqrt{2}$,
$\psi^{(2)}_-=-i(\psi_--\psi^c_+)/\sqrt{2}$, where
$\psi_D=\psi_-+\psi_+$. Setting either $\psi_-$ or $\psi_+$ to zero,
the Weyl spinors $\psi^{(1)}_-$, $\psi^{(2)}_-$ hence also the Majorana
ones $\psi^{(1)}_M$, $\psi^{(2)}_M$ are then no longer independent,
leading back to the construction above.}
\begin{equation}
\psi_M(x)=\psi_-(x)+\psi^c_-(x)\ \ \ ,\ \ \ 
\psi^c_M(x)=\lambda_MC\overline{\psi}^{\rm T}=\psi_M(x)\ ,
\end{equation}
where it is now emphasized that the arbitrary phase factor $\lambda_M$
arising in the definition of spinors which are self-conjugate under charge
conjugation may {\sl a priori\/} be different for each Majorana field.
Consequently, the mode expansion of a Majorana spinor in the helicity
basis is of the form,
\begin{equation}
\psi_M(x)=\int_{(\infty)}\frac{d^3\vec{k}}{(2\pi)^32|\vec{k}|}\,
\sum_{\eta=\pm}\left[e^{-ik\cdot x}u(\vec{k},\eta)a(\vec{k},\eta)+
e^{ik\cdot x}\lambda_M v(\vec{k},\eta)a^\dagger(\vec{k},\eta)\right]\ ,
\end{equation}
where the annihilation and creation operators $a(\vec{k},\eta)$
and $a^\dagger(\vec{k},\eta)$ obey the fermionic algebra
\begin{equation}
\left\{a(\vec{k},\eta),a^\dagger(\vec{k}',\eta')\right\}=
(2\pi)^32|\vec{k}|\delta_{\eta,\eta'}\delta^{(3)}(\vec{k}-\vec{k}')\ .
\end{equation}
In terms of the quanta of the basic Weyl spinor used in the construction,
we thus have the following correspondence\footnote{The complex conjugate
of a complex number $z$ is denoted $z^*$ throughout.},
\begin{equation}
\begin{array}{r c l c r c l}
a(\vec{k},-)&:&b(\vec{k},-)\ \ \ &;&\ \ \ 
a^\dagger(\vec{k},-)&:&b^\dagger(\vec{k},-)\ \ ,\\
a(\vec{k},+)&:&\lambda_M d(\vec{k},+)\ \ \ &;&\ \ \ a^\dagger(\vec{k},+)&:&
\lambda^*_Md^\dagger(\vec{k},+)\ \ ,
\end{array}
\label{eq:correscreation}
\end{equation}
which once again shows that $a^\dagger(\vec{k},\eta)$ is the creation
operator of a particle of momentum $\vec{k}$ and helicity $\eta$, which 
furthermore is in the present case also its own antiparticle.

Note the charge conjugation phase factor
$\lambda_M$ multiplying the creation operator contribution to the mode
expansion of the Majorana quantum field $\psi_M(x)$. This phase factor
corresponds exactly to the ``creation phase factor" whose role has been 
emphasized already in Ref.\cite{Kayser} on different grounds, and again
more recently\cite{Nowa}.

\subsection{Differential cross sections}
\label{Subsect2.2}

All $2\rightarrow 2$ processes of interest in this note are directly
considered in the center-of-mass (CM) frame of the reaction, with a kinematics 
of the form
\begin{equation}
p_1+p_2\rightarrow q_1+q_2\ ,
\end{equation}
the quantities $p_{1,2}$, $q_{1,2}$ standing of course for the four-momenta
of the respective in-coming and out-going massless particles. Given rotation
invariance, and the fact that all particles are of spin 1/2 and of zero mass,
hence of definite helicity, the sole angle of relevance is the CM scattering 
angle $\theta$ between, say, the momenta $\vec{p}_1$ and $\vec{q}_1$. For all 
the reactions listed hereafter, the same order is used for the pairs $(p_1,p_2)$
and $(q_1,q_2)$ of the initial and final particles involved, hence leading 
always to the same interpretation for this angle $\theta$ as being the 
scattering angle between the first particles in these two pairs of in-coming 
and out-going particles.

For specific external particles of definite helicity,
the differential CM cross section of all such processes is given by
\begin{equation}
\frac{d\sigma}{d\Omega_{\hat{q}_1}}=\frac{1}{S_f}\frac{1}{64\pi^2\,s}\,
|{\cal M}|^2\ \ \ ,\ \ \ 
\frac{d\sigma}{d\cos\theta}=\frac{1}{S_f}\frac{1}{32\pi\,s}\,
|{\cal M}|^2\ .
\end{equation}
Here, $\sqrt{s}$ stands for the total invariant energy of the reaction, with
\begin{equation}
s=(p_1+p_2)^2=(q_1+q_2)^2\ ,
\end{equation}
$d\Omega_{\hat{q}_1}$ is the solid angle associated to the outgoing
particle of normalized momentum $\hat{q}_1=\vec{q}_1/|\vec{q}_1|$,
$S_f=2$ or $S_f=1$ depending on whether the two particles---including their
helicity---in the final state are identical or not, respectively, 
and ${\cal M}$ is Feynman's scattering matrix element. Thus, it is only 
through $|{\cal M}|^2$ that the differential cross section depends
on\footnote{Note that this fact implies that relative angular dependencies 
of cross sections are energy independent (within the regime to which the 
four-fermion parametrization applies), while of course reaction rates are 
directly energy dependent with their usual linear dependency in $s$.}
the scattering angle $\theta$. Furthermore, this expression also shows that 
it is sufficient for our purposes to simply determine the amplitude 
${\cal M}$ for each of the relevant processes, a single complex quantity
function of $\theta$. All our results are thus listed in terms of the
amplitude ${\cal M}$ for each process given an arbitrary combination of
helicities for the external states.

\section{The Effective Lagrangian}
\label{Sect3}

Given a choice of external states including their helicities, 
Feynman's amplitude ${\cal M}$ is
determined from the interaction Lagrangian for these particles.
Assuming that the energy $\sqrt{s}$ remains much smaller than any
of the mass scales of the fundamental interactions at work, an effective
four-fermion parametrization of this interaction is warranted, constrained
by the sole requirements of Lorentz invariance and electric charge conservation.
Since fermion number is not necessarily conserved in interactions
involving neutrinos, {\sl a priori\/} one may couple equally well the
neutrino fields and their charge conjugates to the charged fermionic fields.
For the latter, the Dirac fields that will be used represent the usual 
charged leptons (or quarks), rather than their antiparticules.
It is relative to this choice for the charged fields that the neutrino fields
and their charge conjugates are thus specified.

With this understanding in mind, we shall consider all processes involving
neutrinos (or their antineutrinos) of definite flavours $a$ and $b$ as well as
leptons (or their antileptons) of flavours $i$ and $j$, all denoted
as $\nu_a$, $\nu_b$, $\ell^-_i$ and $\ell^-_j$, respectively. The same
notation is used for the associated spinor fields, except for the
indication of the lepton charge. Hence, the total four-fermion
effective Lagrangian that is considered throughout in the case of
Dirac neutrino fields is of the form\footnote{The charge exchange form
of these interactions is used here, but a charge conserving one could
likewise be contemplated, one being related to the other through a Fierz
transformation.}
\begin{equation}
{\cal L}_{\rm eff}=4\frac{g^2}{8M^2}\left[{\cal L}_D+{\cal L}^\dagger_D\right]
\ ,
\end{equation}
where
\begin{equation}
{\cal L}_D={\cal L}_1+{\cal L}_2+{\cal L}_3+{\cal L}_4\ ,
\end{equation}
while each of the separate contributions is given by
\begin{equation}
{\cal L}_1=S_1^{\eta_a,\eta_b}\overline{\nu_a}P_{-\eta_a}\ell_i\
\overline{\ell_j}P_{\eta_b}\nu_b\ +\
V_1^{\eta_a,\eta_b}\overline{\nu_a}\gamma^\mu P_{\eta_a}\ell_i\
\overline{\ell_j}\gamma_\mu P_{\eta_b}\nu_b\ +\
\frac{1}{2}T_1^{\eta_a,\eta_b}\overline{\nu_a}\sigma^{\mu\nu}P_{-\eta_a}\ell_i\
\overline{\ell_j}\sigma_{\mu\nu}P_{\eta_b}\nu_b\ ,
\label{eq:1}
\end{equation}
\begin{equation}
{\cal L}_2=S_2^{\eta_a,\eta_b}\overline{\nu^c_a}P_{\eta_a}\ell_i\
\overline{\ell_j}P_{\eta_b}\nu_b\ +\
V_2^{\eta_a,\eta_b}\overline{\nu^c_a}\gamma^\mu P_{-\eta_a}\ell_i\
\overline{\ell_j}\gamma_\mu P_{\eta_b}\nu_b\ +\
\frac{1}{2}T_2^{\eta_a,\eta_b}\overline{\nu^c_a}\sigma^{\mu\nu}P_{\eta_a}\ell_i\
\overline{\ell_j}\sigma_{\mu\nu}P_{\eta_b}\nu_b\ ,
\end{equation}
\begin{equation}
{\cal L}_3=S_3^{\eta_a,\eta_b}\overline{\nu_a}P_{-\eta_a}\ell_i\
\overline{\ell_j}P_{-\eta_b}\nu^c_b\ +\
V_3^{\eta_a,\eta_b}\overline{\nu_a}\gamma^\mu P_{\eta_a}\ell_i\
\overline{\ell_j}\gamma_\mu P_{-\eta_b}\nu^c_b\ +\
\frac{1}{2}T_3^{\eta_a,\eta_b}\overline{\nu_a}\sigma^{\mu\nu}P_{-\eta_a}\ell_i\
\overline{\ell_j}\sigma_{\mu\nu}P_{-\eta_b}\nu^c_b\ ,
\end{equation}
\begin{equation}
{\cal L}_4=S_4^{\eta_a,\eta_b}\overline{\nu^c_a}P_{\eta_a}\ell_i\
\overline{\ell_j}P_{-\eta_b}\nu^c_b\ +\
V_4^{\eta_a,\eta_b}\overline{\nu^c_a}\gamma^\mu P_{-\eta_a}\ell_i\
\overline{\ell_j}\gamma_\mu P_{-\eta_b}\nu^c_b\ +\
\frac{1}{2}T_4^{\eta_a,\eta_b}\overline{\nu^c_a}\sigma^{\mu\nu}P_{\eta_a}\ell_i\
\overline{\ell_j}\sigma_{\mu\nu}P_{-\eta_b}\nu^c_b\ ,
\end{equation}
an implicit summation over the chiralities $\eta_a$ and $\eta_b$ being
understood of course. On the other hand, it is important to keep in mind
that no summation over the flavour indices $a$ and $b$, nor $i$ and $j$
is implied; all four of these values are fixed at the outset, keeping
open still the possibility that $a$ and $b$ might be equal or not,
and likewise for $i$ and $j$.

The overall normalization factor $4g^2/8M^2$ involves a dimensionless
coupling constant $g$ as well as a mass scale $M$, while the factor $4$
cancels the two $1/2$ factors present in the definition
of the chiral projection operators $P_{\pm\eta_a}$ and $P_{\pm\eta_b}$
which appear in the effective interactions. The motivation for this
choice of normalization is that in the specific limit of the SM, the
effective interaction is normalized precisely in this manner with $g$
then being the SU(2)$_L$ gauge coupling constant $g_L$ and $M$ the massive 
$W^\pm$ gauge boson mass $M_W$, with in particular their tree-level
relation to Fermi's constant, $G_F/\sqrt{2}=g^2_L/(8M^2_W)$ 
(see further details below in the case of the SM).

In the above definitions, the complex coupling coefficients 
$\left\{S,V,T\right\}_{1,2,3,4}^{\eta_a,\eta_b}$ parametrize
the most general four-fermion interactions possible, including the eventuality
of CP violation whenever at least one of these coefficients is complex. 
The choice for the indices $\eta_a$ and $\eta_b$ is made such 
that they each correspond to the chirality and helicity $\eta_a$ or $\eta_b$
of the neutrino spinor fields and associated particles involved in the 
effective coupling, with the chiralities of the leptonic fields being then 
determined according to the selection rules governing scalar, vector or 
tensor couplings, namely these chiralities are those of the associated 
neutrino for vector couplings, and opposite to it for scalar an tensor 
couplings. The situation with regards to tensor couplings is 
particular, in that the relation 
$\sigma^{\mu\nu}\gamma_5=i\epsilon^{\mu\nu\rho\sigma}\sigma_{\rho\sigma}/2$
implies that the couplings $T^{\eta_a,\eta_b}_1$ and $T^{\eta_a,\eta_b}_4$
contribute to the effective interaction only if the chiralities $\eta_a$
and $\eta_b$ are opposite, $\eta_a=-\eta_b$, while the couplings
$T^{\eta_a,\eta_b}_2$ and $T^{\eta_a,\eta_b}_3$ contribute only if
$\eta_a=\eta_b$. These conventions and remarks are also those that apply
to the by now standard four-fermion parametrization used in the 
$\mu-e$ sector\cite{RPP}.

In the case of Majorana neutrino fields, a similar parametrization is
of application, namely 
\begin{equation}
{\cal L}_{\rm eff}=4\frac{g^2}{8M^2}\left[{\cal L}_M+{\cal L}^\dagger_M\right]
\ ,
\end{equation}
where
\begin{equation}
{\cal L}_M=S^{\eta_a,\eta_b}\overline{\nu_a}P_{-\eta_a}\ell_i\
\overline{\ell_j}P_{\eta_b}\nu_b\ +\
V^{\eta_a,\eta_b}\overline{\nu_a}\gamma^\mu P_{\eta_a}\ell_i\
\overline{\ell_j}\gamma_\mu P_{\eta_b}\nu_b\ +\
\frac{1}{2}T^{\eta_a,\eta_b}\overline{\nu_a}\sigma^{\mu\nu}P_{-\eta_a}\ell_i\
\overline{\ell_j}\sigma_{\mu\nu}P_{\eta_b}\nu_b\ .
\label{eq:Maj}
\end{equation}
Compared to the definitions above, and upon using the property 
$\psi^c_M=\psi_M$ characterizing Majorana spinors, we thus have the following 
correspondence between the effective coupling coefficients in the
Majorana and Dirac cases,
\begin{equation}
S^{\eta_a,\eta_b}\ :\ \ \ 
S^{\eta_a,\eta_b}_1+S^{-\eta_a,\eta_b}_2+S^{\eta_a,-\eta_b}_3
+S^{-\eta_a,-\eta_b}_4\ ,
\label{eq:correspondence1}
\end{equation}
\begin{equation}
V^{\eta_a,\eta_b}\ :\ \ \ 
V^{\eta_a,\eta_b}_1+V^{-\eta_a,\eta_b}_2+V^{\eta_a,-\eta_b}_3
+V^{-\eta_a,-\eta_b}_4\ ,
\label{eq:correspondence2}
\end{equation}
\begin{equation}
T^{\eta_a,\eta_b}\ :\ \ \ 
T^{\eta_a,\eta_b}_1+T^{-\eta_a,\eta_b}_2+T^{\eta_a,-\eta_b}_3
+T^{-\eta_a,-\eta_b}_4\ .
\label{eq:correspondence3}
\end{equation}

These effective Lagrangians are still not yet the most general
ones possible, when either $a\ne b$ or $i\ne j$, or both.
In the above, it is implicitly assumed that the flavours
$a$ and $i$, on the one hand, and $b$ and $j$ on the other, couple to one 
another in the ``current$\times$current" representation of these interactions.
One could still add other similar terms in which the roles of
the flavours $a$ and $b$, say, are exchanged, providing still futher
interactions whenever $a\ne b$ or $i\ne j$. Nonetheless, such a possibility
may easily be included in the results hereafter, since the explicit expressions
for the matrix elements ${\cal M}$, rather than the cross sections, which
are provided, are linear in the coupling coefficients.

As a final remark, let us also note that the total neutrino fermionic
number is conserved in these effective interactions only
for couplings of type 1 and 4, $\{S,V,T\}^{\eta_a,\eta_b}_{1,4}$,
whereas those associated to the couplings of type 2 and 3,
$\{S,V,T\}^{\eta_a,\eta_b}_{2,3}$, violate that quantum number by two units.

It is of interest to determine the effective coupling coefficients in the
specific case of the electroweak Standard Model, for which the normalization
factor $4g^2_L/(8M^2_W)$ was discussed previously already. Due to flavour 
conservation rules in that instance, different situations must be 
distinguished, depending on whether only $W^\pm$ or only $Z_0$ exchanges 
are involved, or both. 

Purely $W^\pm$ exchange processes arise when $a=i$ and $b=j$ but also $a\ne b$ 
and $i\ne j$, in which case the only nonvanishing effectif coupling is
\begin{equation}
{\rm SM}:\ \ \ 
a=i\ ;\ b=j\ ;\ a\ne b\ ;\ i\ne j\ :\ \ \ 
V^{-,-}_1=-1\ .
\end{equation}

In the case of purely $Z_0$ neutral current processes, we have for the only
nonvanishing couplings,
\begin{equation}
{\rm SM}:\ \ \ 
(a=b)\ne(i=j)\ :\ \ \ 
S^{-,-}_1=\sin^2\theta_W\ \ \ ,\ \ \ 
V^{-,-}_1=\frac{1}{4}\left(1-2\sin^2\theta_W\right)\ ,
\end{equation}
$\theta_W$ being the usual electroweak gauge mixing angle.
Note that in this situation, the Lagrangians ${\cal L}_D$ and 
${\cal L}^\dagger_D$, or ${\cal L}_M$ and ${\cal L}^\dagger_M$, are
equal.

Finally, charged and neutral exchanges both contribute only when
$a=b=i=j$, in which case the only nonvanishing couplings are
\begin{equation}
{\rm SM}:\ \ \ 
a=b=i=j\ :\ \ \ 
S^{-,-}_1=\sin^2\theta_W\ \ \ ,\ \ \ 
V^{-,-}_1=-\frac{1}{2}+\frac{1}{4}\left(1-2\sin^2\theta_W\right)\ .
\end{equation}
In this case as well, the Lagrangians ${\cal L}_D$ and 
${\cal L}^\dagger_D$, or ${\cal L}_M$ and ${\cal L}^\dagger_M$, are
equal.

Any extra coupling coefficient introduced beyond these ones thus
corresponds to some new physics beyond the Standard Model.
Any particular model beyond the Standard Model predicts specific values
for a subclass of the effective couplings parametrizing the general
expression being used here, to which the general results to be
presented hereafter may thus readily be applied.

The remainder of the calculation proceeds straightforwardly. Given
any choice of external states for the in-coming and out-going particles
with their specific helicities, the substitution of the effective
Lagrangian operator enables the direct evaluation of the associated
matrix element ${\cal M}$ using the Fock algebra of the creation and
annihilation operators that appear in the mode expansions of the fermion
fields. Rather than working out the quantity $|{\cal M}|^2$ through the 
usual trace techniques, it proves much more
efficient to simply substitute for the explicit expressions of the
$u(\vec{k},\eta)$ and $v(\vec{k},\eta)$ spinors solving the free Dirac
equation in the helicity basis and in the chiral representation,
given in Sect.\ref{Subsect2.1}. Choosing a specific CM kinematics configuration
in which only the scattering angle $\theta$ is involved for the reasons
of rotational invariance advocated previously, one then readily
obtains a single complex quantity, namely simply the value for
the amplitude ${\cal M}$ as a function of $\theta$. This is the procedure
that has been applied to each of the processes, leading to the results
listed herafter.

\section{Neutrino Pair Annihilation}
\label{Sect4}

The first general class of processes to be considered is that of
neutrino annihilations into charged lepton (or quark) pairs.
In the Dirac case, these reactions are labelled as follows,

\begin{center}
\underline{$(ab)(ij)$ Dirac neutrino annihilations}

\vspace{10pt}

\begin{tabular}{r c l}
ab1: $\nu_a+\nu_b\rightarrow\ell^-_i+\ell^+_j$&\ \ \ ,\ \ \ &
ab2: $\nu_a+\nu_b\rightarrow\ell^+_i+\ell^-_j$\ \ \ ,\\
ab3: $\nu_a+\bar{\nu}_b\rightarrow\ell^-_i+\ell^+_j$&\ \ \ ,\ \ \ &
ab4: $\nu_a+\bar{\nu}_b\rightarrow\ell^+_i+\ell^-_j$\ \ \ ,\\
ab5: $\bar{\nu}_a+\nu_b\rightarrow\ell^-_i+\ell^+_j$&\ \ \ ,\ \ \ &
ab6: $\bar{\nu}_a+\nu_b\rightarrow\ell^+_i+\ell^-_j$\ \ \ ,\\
ab7: $\bar{\nu}_a+\bar{\nu}_b\rightarrow\ell^-_i+\ell^+_j$&\ \ \ ,\ \ \ &
ab8: $\bar{\nu}_a+\bar{\nu}_b\rightarrow\ell^+_i+\ell^-_j$\ \ \ ,\\
\end{tabular}
\end{center}

\noindent while in the Majorana case, this list reduces to

\begin{center} 
\underline{$(ab)(ij)$ Majorana neutrino annihilations}

\vspace{10pt}

Mab1: $\nu_a+\nu_b\rightarrow\ell^-_i+\ell^+_j$\ \ \ ,\ \ \ 
Mab2: $\nu_a+\nu_b\rightarrow\ell^+_i+\ell^-_j$\ .\\
\end{center}

Due to identical angular momentum selection rules for all these processes, 
the associated matrix element ${\cal M}$ is, for all these ten processes,
of the form
\begin{equation}
\begin{array}{r c c l}
{\cal M}_{(ab)(ij)}&=&-4s\left(\frac{g^2}{8M^2}\right)\,N_1\,\delta_{ij} &
\left\{\delta_{ab}\delta_{\eta_i}^{-\eta_a}\delta_{\eta_j}^{-\eta_b}
\left[A_{11}\sin^2\theta/2+2\delta_{\eta_a,\eta_b}B_{11}(1+\cos^2\theta/2)\right]
\right.\\
 & & & +
\delta_{ab}\delta_{\eta_i}^{\eta_a}\delta_{\eta_j}^{\eta_b}
C_{11}\left[(1+\eta_a\eta_b)-(1-\eta_a\eta_b)\cos^2\theta/2\right] \\
 & & & +
\delta_{\eta_i}^{-\eta_b}\delta_{\eta_j}^{-\eta_a}\eta_a\eta_bD_1
\left[A_{12}\cos^2\theta/2+2\delta_{\eta_a,\eta_b}B_{12}(1+\sin^2\theta/2)\right]
\\
 & & & +
\left.\delta_{\eta_i}^{\eta_b}\delta_{\eta_j}^{\eta_a}\eta_a\eta_b D_1
C_{12}\left[(1+\eta_a\eta_b)-(1-\eta_a\eta_b)\sin^2\theta/2\right]\right\}\\
 & &-4s\left(\frac{g^2}{8M^2}\right)\,N_2\, &
\left\{\delta_{ab}\delta_{\eta_i}^{-\eta_b}\delta_{\eta_j}^{-\eta_a}
\left[A_{21}\cos^2\theta/2+2\delta_{\eta_a,\eta_b}B_{21}(1+\sin^2\theta/2)\right]
\right.\\
 & & & +
\delta_{ab}\delta_{\eta_i}^{\eta_b}\delta_{\eta_j}^{\eta_a}
C_{21}\left[(1+\eta_a\eta_b)-(1-\eta_a\eta_b)\sin^2\theta/2\right] \\
 & & & +
\delta_{\eta_i}^{-\eta_a}\delta_{\eta_j}^{-\eta_b}\eta_a\eta_bD_2
\left[A_{22}\sin^2\theta/2+2\delta_{\eta_a,\eta_b}B_{22}(1+\cos^2\theta/2)\right]
\\
 & & & +
\left.\delta_{\eta_i}^{\eta_a}\delta_{\eta_j}^{\eta_b}\eta_a\eta_b D_2
C_{22}\left[(1+\eta_a\eta_b)-(1-\eta_a\eta_b)\cos^2\theta/2\right]\right\}\ ,
\end{array}
\label{eq:abij}
\end{equation}
where, in agreement with our conventions,
$\theta$ is the scattering angle between the incoming neutrino of
flavour $a$ and the produced charged lepton of flavour $i$.
The particle helicities are $\eta_a$, $\eta_b$, $\eta_i$ and $\eta_j$,
respectively. Table~\ref{Table:abij} lists the values for the constant
phase factors $N_{1,2}$ and $D_{1,2}$ and the subsets of the scalar,
tensor and vector effective couplings constants, in that order, which
define the quantities $A_{11,12,21,22}$, $B_{11,12,21,22}$ and 
$C_{11,12,21,22}$, whether in the case of Dirac of Majorana neutrinos.

The overall phase and sign of this amplitude is of course irrelevant
physically, and is function of the phase convention adopted for the
external $|{\rm In}>$ and $|{\rm Out}>$ states. The latter
were defined by having the associated creation operators acting on the
vacuum state $|0>$ in the same order as that in which the corresponding 
particles are given in the above lists of processes. For example in the 
case of the process ``ab1", we have thus taken
\begin{equation}
|{\rm In}>=b^\dagger_a(\vec{k}_a,\eta_a)\,b^\dagger_b(\vec{k}_b,\eta_b)\,|0>
\ \ \ ,\ \ \ 
|{\rm Out}>=b^\dagger_i(\vec{\ell}_i,\eta_i)\,
d^\dagger_j(\vec{\ell}_j,\eta_j)\,|0>\ \ \ ,
\end{equation}
in a notation that should be self-explanatory. Similarly in the case ``Mab2"
for instance,
\begin{equation}
|{\rm In}>=a^\dagger_a(\vec{k}_a,\eta_a)\,a^\dagger_b(\vec{k}_b,\eta_b)\,|0>
\ \ \ ,\ \ \ 
|{\rm Out}>=d^\dagger_i(\vec{\ell}_i,\eta_i)\,b^\dagger_j(\vec{\ell}_j,\eta_j)\,
|0>\ .
\end{equation}
Obviously, exactly all the same conventions have been used throughout this
work.

\section{Neutrino Pair Production}
\label{Sect5}

Although neutrino pair production processes as such pose a genuine
experimental challenge for their detection, as opposed to processes in which 
they are accompanied for instance by a photon in the final state\cite{Rossi}, 
$\ell^-\ell^+\rightarrow\nu\bar{\nu}\gamma$,
the corresponding list of results is provided here for completeness.
All $2\rightarrow 2$ neutrino pair production processes are labelled 
according to the following list when both neutrinos are of the Dirac type,

\begin{center}

\underline{$(ij)(ab)$ Dirac processes}

\vspace{10pt}

\begin{tabular}{r c l}

ij1: $\ell^-_i+\ell^+_j\rightarrow \nu_a+\nu_b$&\ \ \ ,\ \ \ &
ij2: $\ell^+_i+\ell^-_j\rightarrow \nu_a+\nu_b$\ \ \ ,\\
ij3: $\ell^-_i+\ell^+_j\rightarrow \nu_a+\bar{\nu}_b$&\ \ \ ,\ \ \ &
ij4: $\ell^+_i+\ell^-_j\rightarrow \nu_a+\bar{\nu}_b$\ \ \ ,\\
ij5: $\ell^-_i+\ell^+_j\rightarrow \bar{\nu}_a+\nu_b$&\ \ \ ,\ \ \ &
ij6: $\ell^+_i+\ell^-_j\rightarrow \bar{\nu}_a+\nu_b$\ \ \ ,\\
ij7: $\ell^-_i+\ell^+_j\rightarrow \bar{\nu}_a+\bar{\nu}_b$&\ \ \ ,\ \ \ &
ij8: $\ell^+_i+\ell^-_j\rightarrow \bar{\nu}_a+\bar{\nu}_b$\ \ \ ,\\

\end{tabular}

\end{center}

\noindent while in the Majorana case

\begin{center}
\underline{$(ij)(ab)$ Majorana processes}

\vspace{10pt}

Mij1: $\ell^-_i+\ell^+_j\rightarrow \nu_a+\nu_b$\ \ \ ,\ \ \ 
Mij2: $\ell^+_i+\ell^-_j\rightarrow \nu_a+\nu_b$\ .\\

\end{center}

For all these ten processes, the amplitude ${\cal M}$ is always
of the following form
\begin{equation}
\begin{array}{r c c l}
{\cal M}_{(ij)(ab)}&=&4s\left(\frac{g^2}{8M^2}\right)\,N_1\,\delta_{ij} &
\left\{\delta_{ab}\delta_{\eta_i}^{-\eta_a}\delta_{\eta_j}^{-\eta_b}
\left[A_{11}\sin^2\theta/2+2\delta_{\eta_a,\eta_b}B_{11}(1+\cos^2\theta/2)
\right] \right.\\
 & & & -
\delta_{ab}\delta_{\eta_i}^{\eta_a}\delta_{\eta_j}^{\eta_b}C_{11}
\left[(1+\eta_a\eta_b)-(1-\eta_a\eta_b)\cos^2\theta/2\right] \\
 & & & +
\delta_{\eta_i}^{-\eta_b}\delta_{\eta_j}^{-\eta_a} D_1
\left[A_{12}\cos^2\theta/2+2\delta_{\eta_a,\eta_b}B_{12}(1+\sin^2\theta/2)
\right]
\\
 & & & -
\left.\delta_{\eta_i}^{\eta_b}\delta_{\eta_j}^{\eta_a} D_1 C_{12}
\left[(1+\eta_a\eta_b)-(1-\eta_a\eta_b)\sin^2\theta/2\right]\right\} \\
 & &+4s\left(\frac{g^2}{8M^2}\right)\,N_2\, &
\left\{\delta_{ab}\delta_{\eta_i}^{-\eta_b}\delta_{\eta_j}^{-\eta_a}
\left[A_{21}\cos^2\theta/2+2\delta_{\eta_a,\eta_b}B_{21}(1+\sin^2\theta/2)
\right] \right.\\
 & & & -
\delta_{ab}\delta_{\eta_i}^{\eta_b}\delta_{\eta_j}^{\eta_a} C_{21}
\left[(1+\eta_a\eta_b)-(1-\eta_a\eta_b)\sin^2\theta/2\right] \\
 & & & +
\delta_{\eta_i}^{-\eta_a}\delta_{\eta_j}^{-\eta_b} D_2
\left[A_{22}\sin^2\theta/2+2\delta_{\eta_a,\eta_b}B_{22}(1+\cos^2\theta/2)
\right] \\
 & & & -
\left.\delta_{\eta_i}^{\eta_a}\delta_{\eta_j}^{\eta_b} D_2 C_{22}
\left[(1+\eta_a\eta_b)-(1-\eta_a\eta_b)\cos^2\theta/2\right]\right\}\ ,
\end{array}
\label{eq:ijab}
\end{equation}
with the same conventions as previously, in particular that $\theta$
is the angle between the first lepton of flavour $i$ and the first produced
neutrino of flavour $a$. The different factors and coefficients appearing
in this expression are detailed in Table~\ref{Table:ijab}, whether in
the case of Dirac or Majorana neutrinos.

\section{Neutrino Scattering}
\label{Sect6}

Even though it would suffice in the case of neutrino scattering onto a 
charged lepton to give only two classes of processes, for instance 
$(ai)(bj)$ and $(aj)(bi)$, since the two other classes could be obtained by 
appropriate permutations of indices and of the coupling coefficients
with their complex conjugates, the results for all four classes of processes 
are listed nonetheless, for ease of practical use, and for explicit check 
of expressions through their symmetry properties under such permutations.

\subsection{$(ai)(bj)$ neutrino scattering processes}
\label{Subsect6.1}

In the case of neutrinos of Dirac character, the list of processes is labelled
according to

\begin{center}
\underline{$(ai)(bj)$ Dirac processes}

\vspace{10pt}

\begin{tabular}{r c l}
ai1: $\nu_a+\ell^-_i\rightarrow \nu_b+\ell^-_j$&\ \ \ ,\ \ \ &
ai2: $\nu_a+\ell^+_i\rightarrow \nu_b+\ell^+_j$\ \ \ ,\\
ai3: $\nu_a+\ell^-_i\rightarrow \bar{\nu}_b+\ell^-_j$&\ \ \ ,\ \ \ &
ai4: $\nu_a+\ell^+_i\rightarrow \bar{\nu}_b+\ell^+_j$\ \ \ ,\\
ai5: $\bar{\nu}_a+\ell^-_i\rightarrow \nu_b+\ell^-_j$&\ \ \ ,\ \ \ &
ai6: $\bar{\nu}_a+\ell^+_i\rightarrow \nu_b+\ell^+_j$\ \ \ ,\\
ai7: $\bar{\nu}_a+\ell^-_i\rightarrow \bar{\nu}_b+\ell^-_j$&\ \ \ ,\ \ \ &
ai8: $\bar{\nu}_a+\ell^+_i\rightarrow \bar{\nu}_b+\ell^+_j$\ \ \ ,\\

\end{tabular}
\end{center}

\noindent while in the Majorana case

\begin{center}

\underline{$(ai)(bj)$ Majorana processes}

\vspace{10pt}

Mai1: $\nu_a+\ell^-_i\rightarrow \nu_b+\ell^-_j$\ \ \ ,\ \ \ 
Mai2: $\nu_a+\ell^+_i\rightarrow \nu_b+\ell^+_j$\ .\\

\end{center}

The general amplitude ${\cal M}$ then reads in all ten cases as follows
\begin{equation}
\begin{array}{r c c l}
{\cal M}_{(ai)(bj)}&=&4s\left(\frac{g^2}{8M^2}\right)\,N_1\,\delta_{ij} &
\left\{\delta_{ab}\delta_{\eta_i}^{\eta_a}\delta_{\eta_j}^{\eta_b}
\left[A_{11}-2\delta_{\eta_a,-\eta_b}B_{11}(\cos^2\theta/2-\sin^2\theta/2)\right]
\right.\\
 & & & +
\delta_{ab}\delta_{\eta_i}^{-\eta_a}\delta_{\eta_j}^{-\eta_b}
C_{11}\left[1+\eta_a\eta_b(\cos^2\theta/2-\sin^2\theta/2)\right] \\
 & & & +
\delta_{\eta_i}^{-\eta_b}\delta_{\eta_j}^{-\eta_a}\eta_a\eta_b D_1
\left[A_{12}\cos^2\theta/2-2\delta_{\eta_a,-\eta_b}B_{12}(1+\sin^2\theta/2)\right]
\\
 & & & +
\left.\delta_{\eta_i}^{\eta_b}\delta_{\eta_j}^{\eta_a}\eta_a\eta_b D_1
C_{12}\left[(1+\eta_a\eta_b)-(1-\eta_a\eta_b)\sin^2\theta/2)\right]\right\} \\
 & &+4s\left(\frac{g^2}{8M^2}\right)\,N_2\, &
\left\{\delta_{ab}\delta_{\eta_i}^{-\eta_b}\delta_{\eta_j}^{-\eta_a}
\left[A_{21}\cos^2\theta/2-2\delta_{\eta_a,-\eta_b}B_{21}(1+\sin^2\theta/2)\right]
\right.\\
 & & & +
\delta_{ab}\delta_{\eta_i}^{\eta_b}\delta_{\eta_j}^{\eta_a}
C_{21}\left[(1+\eta_a\eta_b)-(1-\eta_a\eta_b)\sin^2\theta/2\right] \\
 & & & +
\delta_{\eta_i}^{\eta_a}\delta_{\eta_j}^{\eta_b}\eta_a\eta_b D_2 
\left[A_{22}-2\delta_{\eta_a,-\eta_b}B_{22}(\cos^2\theta/2-\sin^2\theta/2)\right]
\\
 & & & +
\left.\delta_{\eta_i}^{-\eta_a}\delta_{\eta_j}^{-\eta_b}\eta_a\eta_b D_2
C_{22}\left[1+\eta_a\eta_b(\cos^2\theta/2-\sin^2\theta/2)\right]\right\}\ ,
\end{array}
\label{eq:aibj}
\end{equation}
$\theta$ being the neutrino scattering angle. The list of factors and
coefficients appearing in this expression is detailed in
Table~\ref{Table:aibj}, both in the Dirac and in the Majorana case.

\subsection{$(aj)(bi)$ neutrino scattering processes}
\label{Subsect6.2}

The list of processes in the Dirac case is labelled according to

\begin{center}
\underline{$(aj)(bi)$ Dirac processes}

\vspace{10pt}

\begin{tabular}{r c l}

aj1: $\nu_a+\ell^-_j\rightarrow \nu_b+\ell^-_i$&\ \ \ ,\ \ \ &
aj2: $\nu_a+\ell^+_j\rightarrow \nu_b+\ell^+_i$\ \ \ ,\\
aj3: $\nu_a+\ell^-_j\rightarrow \bar{\nu}_b+\ell^-_i$&\ \ \ ,\ \ \ &
aj4: $\nu_a+\ell^+_j\rightarrow \bar{\nu}_b+\ell^+_i$\ \ \ ,\\
aj5: $\bar{\nu}_a+\ell^-_j\rightarrow \nu_b+\ell^-_i$&\ \ \ ,\ \ \ &
aj6: $\bar{\nu}_a+\ell^+_j\rightarrow \nu_b+\ell^+_i$\ \ \ ,\\
aj7: $\bar{\nu}_a+\ell^-_j\rightarrow \bar{\nu}_b+\ell^-_i$&\ \ \ ,\ \ \ &
aj8: $\bar{\nu}_a+\ell^+_j\rightarrow \bar{\nu}_b+\ell^+_i$\ \ \ ,\\

\end{tabular}
\end{center}

\noindent while in the Majorana case

\begin{center}

\underline{$(aj)(bi)$ Majorana processes}

\vspace{10pt}

Maj1: $\nu_a+\ell^-_j\rightarrow \nu_b+\ell^-_i$\ \ \ ,\ \ \ 
Maj2: $\nu_a+\ell^+_j\rightarrow \nu_b+\ell^+_i$\ .\\

\end{center}

The general scattering amplitude ${\cal M}$ is of the form
\begin{equation}
\begin{array}{r c c l}
{\cal M}_{(aj)(bi)}&=&4s\left(\frac{g^2}{8M^2}\right)\,N_1\,\delta_{ij} &
\left\{\delta_{ab}\delta_{\eta_i}^{-\eta_a}\delta_{\eta_j}^{-\eta_b}
\left[A_{11}\cos^2\theta/2-2\delta_{\eta_a,-\eta_b}B_{11}(1+\sin^2\theta/2)\right]
\right.\\
 & & & +
\delta_{ab}\delta_{\eta_i}^{\eta_a}\delta_{\eta_j}^{\eta_b}
C_{11}\left[(1+\eta_a\eta_b)-(1-\eta_a\eta_b)\sin^2\theta/2\right] \\
 & & & +
\delta_{\eta_i}^{\eta_b}\delta_{\eta_j}^{\eta_a}\eta_a\eta_b D_1
\left[A_{12}-2\delta_{\eta_a,-\eta_b}B_{12}(\cos^2\theta/2-\sin^2\theta/2)\right]
\\
 & & & +
\left.\delta_{\eta_i}^{-\eta_b}\delta_{\eta_j}^{-\eta_a}\eta_a\eta_b D_1
C_{12}\left[1+\eta_a\eta_b(\cos^2\theta/2-\sin^2\theta/2)\right]\right\} \\
 & &+4s\left(\frac{g^2}{8M^2}\right)\,N_2\, &
\left\{\delta_{ab}\delta_{\eta_i}^{\eta_b}\delta_{\eta_j}^{\eta_a}
\left[A_{21}-2\delta_{\eta_a,-\eta_b}B_{21}(\cos^2\theta/2-\sin^2\theta/2)\right]
\right.\\
 & & & +
\delta_{ab}\delta_{\eta_i}^{-\eta_b}\delta_{\eta_j}^{-\eta_a}
C_{21}\left[1+\eta_a\eta_b(\cos^2\theta/2-\sin^2\theta/2)\right] \\
 & & & +
\delta_{\eta_i}^{-\eta_a}\delta_{\eta_j}^{-\eta_b}\eta_a\eta_b D_2 
\left[A_{22}\cos^2\theta/2-2\delta_{\eta_a,-\eta_b}B_{22}(1+\sin^2\theta/2)\right]
\\
 & & & +
\left.\delta_{\eta_i}^{\eta_a}\delta_{\eta_j}^{\eta_b}\eta_a\eta_b D_2
C_{22}\left[(1+\eta_a\eta_b)-(1-\eta_a\eta_b)\sin^2\theta/2)\right]\right\}\ ,
\end{array}
\label{eq:ajbi}
\end{equation}
the angle $\theta$ being that of the scattered neutrino. Table~\ref{Table:ajbi}
lists the relevant factors and coefficients both in the Dirac and in the
Majorana case.

\subsection{$(bi)(aj)$ neutrino scattering processes}
\label{Subsect6.3}

In the Dirac case, we have the following labelling of processes

\begin{center}

\underline{$(bi)(aj)$ Dirac processes}

\vspace{10pt}

\begin{tabular}{r c l}
bi1: $\nu_b+\ell^-_i\rightarrow \nu_a+\ell^-_j$&\ \ \ ,\ \ \ &
bi2: $\nu_b+\ell^+_i\rightarrow \nu_a+\ell^+_j$\ \ \ ,\\
bi3: $\bar{\nu}_b+\ell^-_i\rightarrow \nu_a+\ell^-_j$&\ \ \ ,\ \ \ &
bi4: $\bar{\nu}_b+\ell^+_i\rightarrow \nu_a+\ell^+_j$\ \ \ ,\\
bi5: $\nu_b+\ell^-_i\rightarrow \bar{\nu}_a+\ell^-_j$&\ \ \ ,\ \ \ &
bi6: $\nu_b+\ell^+_i\rightarrow \bar{\nu}_a+\ell^+_j$\ \ \ ,\\
bi7: $\bar{\nu}_b+\ell^-_i\rightarrow \bar{\nu}_a+\ell^-_j$&\ \ \ ,\ \ \ &
bi8: $\bar{\nu}_b+\ell^+_i\rightarrow \bar{\nu}_a+\ell^+_j$\ \ \ ,\\

\end{tabular}
\end{center}

\noindent while in the Majorana case

\begin{center}

\underline{$(bi)(aj)$ Majorana processes}

\vspace{10pt}

Mbi1: $\nu_b+\ell^-_i\rightarrow \nu_a+\ell^-_j$\ \ \ ,\ \ \ 
Mbi2: $\nu_b+\ell^+_i\rightarrow \nu_a+\ell^+_j$\ .\\

\end{center}

The general scattering amplitude ${\cal M}$ reads
\begin{equation}
\begin{array}{r c c l}
{\cal M}_{(bi)(aj)}&=&4s\left(\frac{g^2}{8M^2}\right)\,N_1\,\delta_{ij} &
\left\{\delta_{ab}\delta_{\eta_i}^{-\eta_a}\delta_{\eta_j}^{-\eta_b}
\left[A_{11}\cos^2\theta/2-2\delta_{\eta_a,-\eta_b}B_{11}(1+\sin^2\theta/2)\right]
\right.\\
 & & & +
\delta_{ab}\delta_{\eta_i}^{\eta_a}\delta_{\eta_j}^{\eta_b}
C_{11}\left[(1+\eta_a\eta_b)-(1-\eta_a\eta_b)\sin^2\theta/2\right] \\
 & & & +
\delta_{\eta_i}^{\eta_b}\delta_{\eta_j}^{\eta_a}\eta_a\eta_b D_1
\left[A_{12}-2\delta_{\eta_a,-\eta_b}B_{12}(\cos^2\theta/2-\sin^2\theta/2)\right]
\\
 & & & +
\left.\delta_{\eta_i}^{-\eta_b}\delta_{\eta_j}^{-\eta_a}\eta_a\eta_b D_1
C_{12}\left[1+\eta_a\eta_b(\cos^2\theta/2-\sin^2\theta/2)\right]\right\} \\
 & &+4s\left(\frac{g^2}{8M^2}\right)\,N_2\, &
\left\{\delta_{ab}\delta_{\eta_i}^{\eta_b}\delta_{\eta_j}^{\eta_a}
\left[A_{21}-2\delta_{\eta_a,-\eta_b}B_{21}(\cos^2\theta/2-\sin^2\theta/2)\right]
\right.\\
 & & & +
\delta_{ab}\delta_{\eta_i}^{-\eta_b}\delta_{\eta_j}^{-\eta_a}
C_{21}\left[1+\eta_a\eta_b(\cos^2\theta/2-\sin^2\theta/2)\right] \\
 & & & +
\delta_{\eta_i}^{-\eta_a}\delta_{\eta_j}^{-\eta_b}\eta_a\eta_b D_2 
\left[A_{22}\cos^2\theta/2-2\delta_{\eta_a,-\eta_b}B_{22}(1+\sin^2\theta/2)\right]
\\
 & & & +
\left.\delta_{\eta_i}^{\eta_a}\delta_{\eta_j}^{\eta_b}\eta_a\eta_b D_2
C_{22}\left[(1+\eta_a\eta_b)-(1-\eta_a\eta_b)\sin^2\theta/2)\right]\right\}\ ,
\end{array}
\label{eq:biaj}
\end{equation}
$\theta$ being of course the neutrino scattering angle. The factors and
coefficients appearing in this representation are detailed in
Table~\ref{Table:biaj}.

\subsection{$(bj)(ai)$ neutrino scattering processes}
\label{Subsect6.4}

Processes in the Dirac case are labelled according to

\begin{center}

\underline{$(bj)(ai)$ Dirac processes}

\vspace{10pt}

\begin{tabular}{r c l}
bj1: $\nu_b+\ell^-_j\rightarrow \nu_a+\ell^-_i$&\ \ \ ,\ \ \ &
bj2: $\nu_b+\ell^+_j\rightarrow \nu_a+\ell^+_i$\ \ \ ,\\
bj3: $\bar{\nu}_b+\ell^-_j\rightarrow \nu_a+\ell^-_i$&\ \ \ ,\ \ \ &
bj4: $\bar{\nu}_b+\ell^+_j\rightarrow \nu_a+\ell^+_i$\ \ \ ,\\
bj5: $\nu_b+\ell^-_j\rightarrow \bar{\nu}_a+\ell^-_i$&\ \ \ ,\ \ \ &
bj6: $\nu_b+\ell^+_j\rightarrow \bar{\nu}_a+\ell^+_i$\ \ \ ,\\
bj7: $\bar{\nu}_b+\ell^-_j\rightarrow \bar{\nu}_a+\ell^-_i$&\ \ \ ,\ \ \ &
bj8: $\bar{\nu}_b+\ell^+_j\rightarrow \bar{\nu}_a+\ell^+_i$\ \ \ ,\\

\end{tabular}
\end{center}

\noindent while in the Majorana case

\begin{center}

\underline{$(bj)(ai)$ Majorana processes}

\vspace{10pt}

Mbj1: $\nu_b+\ell^-_j\rightarrow \nu_a+\ell^-_i$\ \ \ ,\ \ \
Mbj2: $\nu_b+\ell^+_j\rightarrow \nu_a+\ell^+_i$\ .\\

\end{center}

The general scattering amplitude ${\cal M}$ is given by
\begin{equation}
\begin{array}{r c c l}
{\cal M}_{(bj)(ai)}&=&4s\left(\frac{g^2}{8M^2}\right)\,N_1\,\delta_{ij} &
\left\{\delta_{ab}\delta_{\eta_i}^{\eta_a}\delta_{\eta_j}^{\eta_b}
\left[A_{11}-2\delta_{\eta_a,-\eta_b}B_{11}(\cos^2\theta/2-\sin^2\theta/2)\right]
\right.\\
 & & & +
\delta_{ab}\delta_{\eta_i}^{-\eta_a}\delta_{\eta_j}^{-\eta_b}
C_{11}\left[1+\eta_a\eta_b(\cos^2\theta/2-\sin^2\theta/2)\right] \\
 & & & +
\delta_{\eta_i}^{-\eta_b}\delta_{\eta_j}^{-\eta_a}\eta_a\eta_b D_1
\left[A_{12}\cos^2\theta/2-2\delta_{\eta_a,-\eta_b}B_{12}(1+\sin^2\theta/2)\right]
\\
 & & & +
\left.\delta_{\eta_i}^{\eta_b}\delta_{\eta_j}^{\eta_a}\eta_a\eta_b D_1
C_{12}\left[(1+\eta_a\eta_b)-(1-\eta_a\eta_b)\sin^2\theta/2)\right]\right\} \\
 & &+4s\left(\frac{g^2}{8M^2}\right)\,N_2\, &
\left\{\delta_{ab}\delta_{\eta_i}^{-\eta_b}\delta_{\eta_j}^{-\eta_a}
\left[A_{21}\cos^2\theta/2-2\delta_{\eta_a,-\eta_b}B_{21}(1+\sin^2\theta/2)\right]
\right.\\
 & & & +
\delta_{ab}\delta_{\eta_i}^{\eta_b}\delta_{\eta_j}^{\eta_a}
C_{21}\left[(1+\eta_a\eta_b)-(1-\eta_a\eta_b)\sin^2\theta/2\right] \\
 & & & +
\delta_{\eta_i}^{\eta_a}\delta_{\eta_j}^{\eta_b}\eta_a\eta_b D_2 
\left[A_{22}-2\delta_{\eta_a,-\eta_b}B_{22}(\cos^2\theta/2-\sin^2\theta/2)\right]
\\
 & & & +
\left.\delta_{\eta_i}^{-\eta_a}\delta_{\eta_j}^{-\eta_b}\eta_a\eta_b D_2
C_{22}\left[1+\eta_a\eta_b(\cos^2\theta/2-\sin^2\theta/2)\right]\right\}\ ,
\end{array}
\label{eq:bjai}
\end{equation}
with $\theta$ being the neutrino scattering angle. Table~\ref{Table:bjai}
lists the relevant factors and coefficients both in the Dirac and Majorana
cases.

\section{Exploratory Examples}
\label{Sect7}

Before turning to some simple illustrative examples of the potential
physics reach of these $2\rightarrow 2$ processes involving neutrinos,
let us point out the following simple fact. For each of the six classes
of ten processes above, when comparing the cases with Dirac or with
Majorana neutrinos, one notices that Dirac neutrino processes labelled 
``xy$n$" (``x" and ``y" each being one of the neutrino or lepton flavour 
symbols and $n$ being an integer) must be in correspondence with the 
Majorana process ``Mxy1" for $n=1,3,5,7$, and with the
Majorana process ``Mxy2" for $n=2,4,6,8$. Namely, given the
correspondence (\ref{eq:correscreation}), and for a specific choice of 
the external particle helicities, in each case the sum of the
corresponding four Dirac amplitudes ${\cal M}$ must coincide with the
amplitude ${\cal M}$ for the Majorana process through the association
of coupling coefficients described in 
(\ref{eq:correspondence1})-(\ref{eq:correspondence3}). It is
straightforward to check that this correspondence is indeed obtained.

This fact, with in particular an identical angular parametrization
of the differential cross section for all ten processes belonging to
each one of the above six $2\rightarrow 2$ classes, also implies that
a model independent discrimination between the Dirac of Majorana character
of neutrinos would require stringent precision requirements for systematic
measurements in which different helicity combinations are compared,
in order to isolate the relevant coefficients $A_{\alpha\beta}$,
$B_{\alpha\beta}$ and $C_{\alpha\beta}$ ($\alpha,\beta=1,2$), and
eventually determine their possible relations as applicable either to 
the Dirac or to the Majorana case.

However, the four-fermion parametrization used is far more general than
what is usually achieved in any specific model beyond the SM, and as a
general rule, couplings of type 2, 3 and 4, namely
$S^{\eta_a,\eta_b}_{2,3,4}$, $V^{\eta_a,\eta_b}_{2,3,4}$ and
$T^{\eta_a,\eta_b}_{2,3,4}$, do not arise. Under such a situation
which remains quite model independent, it is clear that in principle
there exist large classes of processes which, simply through the
angular dependency of their differential cross sections, should enable
the experimental discrimination between the Dirac or Majorana neutrino
character. However, without any prior knowledge of the order of magnitude
of the coupling coefficients which determine the relative
strengths of different angular dependencies, such a discrimination in
a model independent manner requires that at least one of the flavour pairs
$(ab)$ or $(ij)$ be identical. Indeed, one gets different
angular contributions to the Dirac or Majorana amplitudes
${\cal M}$ provided only at least one of the set of terms proportional
either to $\delta_{ij}$ or to $\delta_{ab}$, or both, contributes to the
differential cross section. Nevertheless, this still leaves open
{\sl a priori\/} quite some possibilities, contingent onto the properties
of the neutrino beams that would become available in the future, in 
particular their flavour and helicity contents. 

In fact, in the latter respect, any helicity content of a neutrino beam
other than left-handed for what is thought to be a neutrino and
right-handed for what is thought to an antineutrino in the Dirac case,
depends on possible interactions beyond the SM that might contribute to
the neutrino beam production mechanism. In any event, such a helicity
``contamination" of a beam must be expected not to exceed, say, one percent,
given present limits on neutrino helicities\cite{RPP}. Depending on
the intensities of beams to become available at neutrino factories, 
this might provide an additional aspect of physical interest nonetheless. 
For the time being however, let us conservatively assume that available 
beams would only be purely left-handed for would-be Dirac neutrinos and 
purely right-handed for would-be Dirac antineutrinos.

The experimental possibility to eventually discriminate through neutrino
annihilation and scattering processes between their Dirac or Majorana
character using the difference in the angular dependency of the 
associated cross sections, is also contingent on the strength of any
new interaction beyond those of the SM---as is also the neutrinoless
double $\beta$-decay process for that matter\cite{2nu0}---, since in the 
latter model such a possibility simply evaporates in the massless 
limit\cite{Confusion}. Hence, even though 
the possibility exists in principle, its actual experimental realization 
hinges, on the one hand, on sufficiently intense beams at neutrino factories 
to allow for reasonably precise angular measurements, and on the other hand, 
on the physical existence of a new interaction different from $(V-A)$ 
that couples sufficiently strongly to neutrinos as compared to those of the SM. 
Clearly, a definite assessment of the physics potential of such an approach 
to the Dirac-Majorana neutrino issue requires a systematic and dedicated 
analysis which is not attempted here, based on actual neutrino factory 
designs as presently foreseen, and the general low-energy parametrization 
developed here.

Besides the potential resolution of the Dirac-Majorana neutrino issue,
intense neutrino beams should also help turn into reality the systematic
determination of the neutrino electroweak interactions, in a manner
similar to what has been done in the leptonic $(\nu_\mu\mu)(e\nu_e)$
and semi-leptonic $(ud)(e\nu_e)$ sectors\cite{DQ,RPP}. Through detailed
precisions measurements in different combinations of flavour and
helicity channels, which is also part of the neutrino oscillation
programmes at neutrino factories, it should become
possible to set ever more stringent experimental bounds on the different
coupling constants that parametrize the general effective four-fermion
interaction, and search for a lack of overlap with those of the SM. In the 
same way that Refs.\cite{JTW,WF,RPP} provide the expressions of all possible 
observables in terms of such coupling coefficients for the above leptonic and 
semi-leptonic channels, the results listed in this note
provide those for all $2\rightarrow 2$ processes in which two neutrinos and
two charged fermions take part, whatever their mass eigenstates.

In the remainder of this section, simple illustrative examples are briefly
considered, whose sole purpose is a first exploration of some of the above 
issues. The emphasis here is only on the Dirac-Majorana neutrino issue.

First, let us consider the elastic scattering\footnote{The $\nu_\mu$ component
is indeed dominant in neutrino beams.}
\begin{equation}
\nu_\mu+e^-\rightarrow\nu_\mu+e^-\ ,
\end{equation}
which is thus a reaction of type ``ai1" belonging to the $(ai)(bj)$ class,
with $a=b\ne i=j$. Having in mind for instance the
left-right symmetric extensions\cite{Moha} of the SM, based on the gauge group 
$SU(2)_L\times SU(2)_R\times U(1)_{B-L}$, let us assume for utmost simplicity
that besides the $S^{-,-}_1$ and $V^{-,-}_1$ couplings of the SM, the only
nonvanishing extra couplings are $S^{+,+}_1$ and $V^{+,+}_1$. In the
simplest minded situation where the two chiral sectors are identical in
as far as is possible in all their aspects and do not mix, we thus have,
\begin{equation}
S^{-,-}_1=\sin^2\theta_W\ \ ,\ \ 
V^{-,-}_1=\frac{1}{4}[1-2\sin^2\theta_W]\ \ ,\ \ 
S^{+,+}_1=\delta\sin^2\theta_W\ \ ,\ \ 
V^{+,+}_1=\delta\frac{1}{4}[1-2\sin^2\theta_W]\ ,
\end{equation}
where
\begin{equation}
\delta=\frac{M^2_1}{M^2_2}\ ,
\end{equation}
is the physical light $W^\pm_1$ to heavy $W^\pm_2$ squared gauge boson
masses ratio, with $M_1\simeq 80$ GeV and $M_2>720$~GeV\cite{RPP},
and $\theta_W$ is the usual electroweak gauge mixing angle, 
$\sin^2\theta_W\simeq 0.231$\cite{RPP}.
Assuming then that the in-coming $\nu_\mu$ neutrino is necessarily
left-handed, only two of the eight possible scattering amplitudes
${\cal M}$ are nonvanishing due to helicity selection rules, namely those with
$(\eta_a,\eta_i;\eta_b,\eta_j)=(-,-;-,-), (-,+;-,+)$. Since the
angular dependency of each of these two amplitudes is identical whether
the neutrinos are Dirac or Majorana, the sole difference being in
their absolute normalization, and given the difficulty in performing
absolute cross section measurements, let us consider the summation over
all processes irrespective of the particle helicities, except of course
for that of the in-coming $\nu_\mu$, thus corresponding to an unpolarized
measurement. In the Dirac case, one then finds
\begin{equation}
|{\cal M}_{-,-;-,-}|^2+|{\cal M}_{-,+;-,+}|^2=(4s)^2
\left(\frac{g^2}{8M^2}\right)^2\left[4{\rm Re}\,V^{-,-}_1\right]^2\,
\left\{1+\left[\frac{{\rm Re}\,S^{-,-}_1}{4{\rm Re}\,V^{-,-}_1}\right]^2\,
(1+\cos\theta)^2\right\}\ ,
\end{equation}
while in the Majorana case,
\begin{equation}
\begin{array}{r l}
|{\cal M}_{-,-;-,-}|^2+|{\cal M}_{-,+;-,+}|^2=&(4s)^2
\left(\frac{g^2}{8M^2}\right)^2\left[4{\rm Re}\,V^{-,-}+
2{\rm Re}\,S^{+,+}\right]^2\,\\
&\times \left\{1+\left[\frac{{\rm Re}\,S^{-,-}+2{\rm Re}\,V^{+,+}}
{4{\rm Re}\,V^{-,-}+2{\rm Re}\,S^{+,+}}\right]^2\,
(1+\cos\theta)^2\right\}\ .
\end{array}
\end{equation}
Consequently, if there are indeed interactions whose chirality structure
is different from those of the SM, processes for Dirac or Majorana neutrinos
do possess different angular properties, enabling in principle the
discrimination between the two cases through precision measurements
of the angular dependency of the cross section, in the present case by
comparing the strength of the $(1+\cos\theta)^2$ term to its value predicted
in the SM. Unfortunately, in this specific case and under the very
restrictive form of the coefficients $S^{+,+}_1$ and $V^{+,+}_1$ considered
above, limits on the possible extra interaction are already such that it 
is too weak to render any deviation observable, the relative variation in 
the relevant factor being less than a percent given the small value for 
$\delta\le(80\ {\rm GeV}/720\ {\rm GeV})^2 \simeq 1.23\times 10^{-2}$.

Scalar or tensor couplings being typically less well constrained than 
vector ones, let us now consider the possibility of an extra scalar
interaction, for either of the following two elastic scattering reactions,
\begin{equation}
\nu_\mu+e^-\rightarrow\nu_\mu+e^-\ \ \ ,\ \ \ 
\nu_\mu+\mu^-\rightarrow\nu_\mu+\mu^-\ .
\end{equation}
Both these reactions are of the ``ai1" type in the $(ai)(bj)$ class,
with $a=b\ne i=j$ in the first case, and $a=b=i=j$ in the second.
Within the SM, the corresponding nonvanishing couplings are thus
\begin{equation}
S^{-,-}_1=\sin^2\theta_W\ \ \ ,\ \ \ 
V^{-,-}_1=\pm\frac{1}{4}\left[1\mp 2\sin^2\theta_W\right]\ ,
\end{equation}
where the upper (resp. lower) sign is for the $(\nu_\mu e)$
(resp. $(\nu_\mu\mu)$) reaction. Assuming that $S^{+,-}_1$ is the
sole nonvanishing extra interaction, and given a left-handed in-coming 
$\nu_\mu$, one finds that whether in the Dirac or Majorana case the only 
nonvanishing amplitudes ${\cal M}$ correspond to the
following helicity combinations: $(\eta_a,\eta_i;\eta_b,\eta_j)=
(-,-;-,-),(-,-;+,+),(-,+;-,+)$. Considering again the situation of 
an unpolarized measurement, the sum over these polarization states reads,
\begin{equation}
\sum|{\cal M}|^2=(4s)^2\left(\frac{g^2}{8M^2}\right)^2\,
\left\{\left[4{\rm Re}\,V^{-,-}_1\right]^2+
\left[{\rm Re}\,S^{-,-}_1\right]^2(1+\cos\theta)^2+
\frac{1}{4}|S^{+,-}_1|^2(1\pm\cos\theta)^2\right\}\ ,
\end{equation}
where in the last term the upper sign corresponds to the Dirac case,
and the lower sign to the Majorana case. Thus once again, we see that
any new interaction whose chirality structure differs from those of the SM
leads to processes in which the angular dependency discriminates
between Dirac and Majorana neutrinos. Taking as an illustration a value
$|S^{+,-}_1|=0.10$ which is a typical upper-bound on such a coupling in 
the leptonic $(e\mu)$ sector\cite{RPP}, one finds a 10\% sensitivity in 
the forward-backward asymmetry. A reasonably precise measurement of the 
differential cross section, and in particular a fit to the expected 
distributions in either case, thus offers the prospect to resolve the 
Dirac-Majorana neutrino issue at neutrino factories. A dedicated study 
should hopefully confirm the present exploratory assessment.

To also highlight the potentiel interest of intersecting neutrino beams,
as a final example let us consider the following two annihilation reactions,
\begin{equation}
\nu_\mu+\nu_\mu\rightarrow e^-+e^+\ \ \ ,\ \ \ 
\nu_\mu+\nu_\mu\rightarrow \mu^-+\mu^+\ .
\end{equation}
These two processes are of the type ``ab1" in the $(ab)(ij)$ class,
with $a=b\ne i=j$ in the first case, and $a=b=i=j$ in the second case,
and thus again with the following interactions in the SM,
\begin{equation}
S^{-,-}_1=\sin^2\theta_W\ \ \ ,\ \ \ 
V^{-,-}_1=\pm\frac{1}{4}\left[1\mp 2\sin^2\theta_W\right]\ ,
\end{equation}
where the upper (resp. lower) sign is associated to the first
(resp. second) reaction. Assuming now that none of the other possible
coupling coefficients belongs to the classes $S^{\eta_a,\eta_b}_{2,3,4}$,
$V^{\eta_a,\eta_b}_{2,3,4}$ and $T^{\eta_a,\eta_b}_{2,3,4}$, one readily
finds that given left-handed initial $\nu_\mu$ neutrinos only, the
amplitude ${\cal M}$ for these processes vanishes identically in the
Dirac case, but not in the Majorana case, with then a specific angular
dependency in the latter case which is function of the interactions that 
might contribute. Even though the detection of the final state products 
should be straightforward, the difficulty lies of course in the density of the
initial neutrino beams even for very intense ones, implying thus an extremely
low rate. Nonetheless, even if only through a single event,
the observation of either of the above processes, or similar ones
for other neutrino mass eigenstates, would definitely help settle the
Dirac-Majorana neutrino puzzle through accelerator experiments.

\section{Conclusions}
\label{Sect8}
 
In the present work, all possible $2\rightarrow 2$ processes involving
definite mass eigenstates of two neutrinos and two charged fermions have
been considered in the massless limit, on the basis of the most general
four-fermion effective Lagrangian possible. All interactions, whether
preserving the neutrino fermion number or not, and for whatever
helicities of the external particles, have been included. The Feynman
amplitudes in the center-of-mass frame have been listed for all these 
processes, from which the relevant differential cross sections readily follow.
Since any particular model for physics beyond the Standard Model predicts
specific values for the four-fermion nonderivative effective coupling 
coefficients, this analysis should be of value to assess the low energy
merits of any such model. 

In the same way as has been done for $\beta$- and $\mu$-decay and
$\mu$-capture through analogous four-fermion effective 
parametrizations\cite{RPP,JG},
the advent in the foreseeable future of neutrino factories with their
intense beams is the main motivation for the considerations developed here.
Of direct interest is the systematic study of the electroweak interactions
in the neutrino sector, by setting ever more stringent limits on the
effective interactions of these particles through precision measurements.
Another physics issue of great topical interest that could be addressed
through such experiments is that of the Dirac-Majorana discrimination of
the character of neutrinos. In agreement with the Dirac-Majorana confusion
theorem\cite{Confusion}, as soon as interactions with a chirality structure 
different from the $(V-A)$ one of the Standard Model are introduced, there 
exist processes which in principle distinguish between these two 
possible characters of the neutrino through the angular dependency of 
differential cross sections, even in the massless limit. The sensitivity of 
such reactions however, is contingent of course on the relative strength of 
these new interactions beyond the Standard Model. Nonetheless, some simple
examples of such a situation were briefly described, albeit not following
any systematic investigation.

The main purpose of this work is to provide the general results for
the Feynman amplitudes for all possible $2\rightarrow 2$ processes with
two neutrinos. On that basis, it should now be possible to develop a detailed 
and dedicated analysis of the potential reach of different such reactions 
towards the above physics issues, given a specific design both of neutrino 
beams and their intensities, and of detector set-ups. Besides the great 
interest to be found in neutrino scattering experiments, the possibilities 
offered by intersecting neutrino beams should not be dismissed offhand without 
first a dedicated assessment as well, the more so since they could possibly 
run in parasitic mode in conjunction with other experiments given the proper
neutrino beam geometrical layout.

\vspace{20pt}

\noindent{\bf Acknowledgements}

JM acknowledges the financial support of the ``Coop\'eration
Universitaire au D\'evelop\-pe\-ment, Commission Interuniversitaire
Francophone" (CUD-CIUF) of the Belgian French Speaking Community, and wishes
to thank the Institute of Nuclear Physics (Catholic University
of Louvain, Belgium) for its hospitality while this work was being
pursued. JG wishes to thank the C.N. Yang Institute for Theoretical Physics 
(State University of New York at Stony Brook, USA) for its hospitality
during the Summer 2001 while part of this work was completed.

\clearpage

\setlength{\textwidth}{200mm}
\setlength{\textheight}{225mm}
\setlength{\topmargin}{-5mm}
\setlength{\oddsidemargin}{-20mm}
\setlength{\evensidemargin}{-20mm}

\begin{table}
\begin{center}
\begin{tabular}{|c||c|c|c|c|c|c|c|c||c|c|}
\hline
 & ab1 & ab2 & ab3 & ab4 & ab5 & ab6 & ab7 & ab8 & Mab1 & Mab2 \\
\hline\hline
$N_1$ & 
$\eta_a\lambda_a^*$ & $\eta_a\lambda_b^*$ & 
$\eta_a$ & $\eta_a$ & $\eta_a$ & $\eta_a$ &
$\eta_a\lambda_b$ & $\eta_a\lambda_a$ & 
$\eta_a\lambda_a^*$ & $\eta_a\lambda_b^*$ \\
\hline\hline
$A_{11}$ &
$S_2^{\eta_b,\eta_a}$ & $S_3^{\eta_b,\eta_a*}$ &
$S_1^{-\eta_b,\eta_a}$ & $S_4^{-\eta_b,\eta_a*}$ &
$S_4^{\eta_b,-\eta_a}$ & $S_1^{\eta_b,-\eta_a*}$ &
$S_3^{-\eta_b,-\eta_a}$ & $S_2^{-\eta_b,-\eta_a*}$ &
$S^{-\eta_b,\eta_a}$ & $S^{\eta_b,-\eta_a*}$ \\
\hline
$B_{11}$ &
$T_2^{\eta_b,\eta_a}$ & $T_3^{\eta_b,\eta_a*}$ &
$T_1^{-\eta_b,\eta_a}$ & $T_4^{-\eta_b,\eta_a*}$ &
$T_4^{\eta_b,-\eta_a}$ & $T_1^{\eta_b,-\eta_a*}$ &
$T_3^{-\eta_b,-\eta_a}$ & $T_2^{-\eta_b,-\eta_a*}$ &
$T^{-\eta_b,\eta_a}$ & $T^{\eta_b,-\eta_a*}$ \\
\hline
$C_{11}$ &
$V_2^{\eta_b,\eta_a}$ & $V_3^{\eta_b,\eta_a*}$ &
$V_1^{-\eta_b,\eta_a}$ & $V_4^{-\eta_b,\eta_a*}$ &
$V_4^{\eta_b,-\eta_a}$ & $V_1^{\eta_b,-\eta_a*}$ &
$V_3^{-\eta_b,-\eta_a}$ & $V_2^{-\eta_b,-\eta_a*}$ &
$V^{-\eta_b,\eta_a}$ & $V^{\eta_b,-\eta_a*}$ \\
\hline\hline
$D_1$ &
$1$ & $1$ & $\lambda_a^*\lambda_b$ & $1$ & $1$ &
$\lambda_a\lambda_b^*$ & $1$ & $1$ & $1$ & $1$ \\
\hline
$A_{12}$ &
$S_2^{\eta_a,\eta_b}$ & $S_3^{\eta_a,\eta_b*}$ &
$S_4^{\eta_a,-\eta_b}$ & $S_1^{\eta_a,-\eta_b*}$ &
$S_1^{-\eta_a,\eta_b}$ & $S_4^{-\eta_a,\eta_b*}$ &
$S_3^{-\eta_a,-\eta_b}$ & $S_2^{-\eta_a,-\eta_b*}$ &
$S^{-\eta_a,\eta_b}$ & $S^{\eta_a,-\eta_b*}$ \\
\hline
$B_{12}$ &
$T_2^{\eta_a,\eta_b}$ & $T_3^{\eta_a,\eta_b*}$ &
$T_4^{\eta_a,-\eta_b}$ & $T_1^{\eta_a,-\eta_b*}$ &
$T_1^{-\eta_a,\eta_b}$ & $T_4^{-\eta_a,\eta_b*}$ &
$T_3^{-\eta_a,-\eta_b}$ & $T_2^{-\eta_a,-\eta_b*}$ &
$T^{-\eta_a,\eta_b}$ & $T^{\eta_a,-\eta_b*}$ \\
\hline
$C_{12}$ &
$V_2^{\eta_a,\eta_b}$ & $V_3^{\eta_a,\eta_b*}$ &
$V_4^{\eta_a,-\eta_b}$ & $V_1^{\eta_a,-\eta_b*}$ &
$V_1^{-\eta_a,\eta_b}$ & $V_4^{-\eta_a,\eta_b*}$ &
$V_3^{-\eta_a,-\eta_b}$ & $V_2^{-\eta_a,-\eta_b*}$ &
$V^{-\eta_a,\eta_b}$ & $V^{\eta_a,-\eta_b*}$ \\
\hline\hline\hline
$N_2$ & 
$\eta_b\lambda_b^*$ & $\eta_b\lambda_a^*$ & 
$\eta_b$ & $\eta_b$ & $\eta_b$ & $\eta_b$ &
$\eta_b\lambda_a$ & $\eta_b\lambda_b$ & 
$\eta_b\lambda_b^*$ & $\eta_b\lambda_a^*$ \\
\hline\hline
$A_{21}$ &
$S_3^{\eta_b,\eta_a*}$ & $S_2^{\eta_b,\eta_a}$ &
$S_4^{-\eta_b,\eta_a*}$ & $S_1^{-\eta_b,\eta_a}$ &
$S_1^{\eta_b,-\eta_a*}$ & $S_4^{\eta_b,-\eta_a}$ &
$S_2^{-\eta_b,-\eta_a*}$ & $S_3^{-\eta_b,-\eta_a}$ &
$S^{\eta_b,-\eta_a*}$ & $S^{-\eta_b,\eta_a}$ \\
\hline
$B_{21}$ &
$T_3^{\eta_b,\eta_a*}$ & $T_2^{\eta_b,\eta_a}$ &
$T_4^{-\eta_b,\eta_a*}$ & $T_1^{-\eta_b,\eta_a}$ &
$T_1^{\eta_b,-\eta_a*}$ & $T_4^{\eta_b,-\eta_a}$ &
$T_2^{-\eta_b,-\eta_a*}$ & $T_3^{-\eta_b,-\eta_a}$ &
$T^{\eta_b,-\eta_a*}$ & $T^{-\eta_b,\eta_a}$ \\
\hline
$C_{21}$ &
$V_3^{\eta_b,\eta_a*}$ & $V_2^{\eta_b,\eta_a}$ &
$V_4^{-\eta_b,\eta_a*}$ & $V_1^{-\eta_b,\eta_a}$ &
$V_1^{\eta_b,-\eta_a*}$ & $V_4^{\eta_b,-\eta_a}$ &
$V_2^{-\eta_b,-\eta_a*}$ & $V_3^{-\eta_b,-\eta_a}$ &
$V^{\eta_b,-\eta_a*}$ & $V^{-\eta_b,\eta_a}$ \\
\hline\hline
$D_2$ &
$1$ & $1$ & $1$ & $\lambda_a^*\lambda_b$ & $\lambda_a\lambda_b^*$ &
$1$ & $1$ & $1$ & $1$ & $1$ \\
\hline
$A_{22}$ &
$S_3^{\eta_a,\eta_b*}$ & $S_2^{\eta_a,\eta_b}$ &
$S_1^{\eta_a,-\eta_b*}$ & $S_4^{\eta_a,-\eta_b}$ &
$S_4^{-\eta_a,\eta_b*}$ & $S_1^{-\eta_a,\eta_b}$ &
$S_2^{-\eta_a,-\eta_b*}$ & $S_3^{-\eta_a,-\eta_b}$ &
$S^{\eta_a,-\eta_b*}$ & $S^{-\eta_a,\eta_b}$ \\
\hline
$B_{22}$ &
$T_3^{\eta_a,\eta_b*}$ & $T_2^{\eta_a,\eta_b}$ &
$T_1^{\eta_a,-\eta_b*}$ & $T_4^{\eta_a,-\eta_b}$ &
$T_4^{-\eta_a,\eta_b*}$ & $T_1^{-\eta_a,\eta_b}$ &
$T_2^{-\eta_a,-\eta_b*}$ & $T_3^{-\eta_a,-\eta_b}$ &
$T^{\eta_a,-\eta_b*}$ & $T^{-\eta_a,\eta_b}$ \\
\hline
$C_{22}$ &
$V_3^{\eta_a,\eta_b*}$ & $V_2^{\eta_a,\eta_b}$ &
$V_1^{\eta_a,-\eta_b*}$ & $V_4^{\eta_a,-\eta_b}$ &
$V_4^{-\eta_a,\eta_b*}$ & $V_1^{-\eta_a,\eta_b}$ &
$V_2^{-\eta_a,-\eta_b*}$ & $V_3^{-\eta_a,-\eta_b}$ &
$V^{\eta_a,-\eta_b*}$ & $V^{-\eta_a,\eta_b}$ \\
\hline
\end{tabular}
\caption[]{List of the constant factors appearing in the amplitude 
(\ref{eq:abij}) for all $(ab)(ij)$ neutrino annihilation processes according 
to their labelling defined in Sect.\ref{Sect4}.}
\label{Table:abij}
\end{center}
\end{table}

\vspace{10pt}

\begin{table}
\begin{center}
\begin{tabular}{|c||c|c|c|c|c|c|c|c||c|c|}
\hline
 & ij1 & ij2 & ij3 & ij4 & ij5 & ij6 & ij7 & ij8 & Mij1 & Mij2 \\
\hline\hline
$N_1$ & 
$\eta_a\lambda_a$ & $\eta_a\lambda_b$ & 
$\eta_a$ & $\eta_a$ & $\eta_a$ & $\eta_a$ &
$\eta_a\lambda_b^*$ & $\eta_a\lambda_a^*$ & 
$\eta_a\lambda_a$ & $\eta_a\lambda_b$ \\
\hline\hline
$A_{11}$ &
$S_2^{\eta_b,\eta_a*}$ & $S_3^{\eta_b,\eta_a}$ &
$S_1^{-\eta_b,\eta_a*}$ & $S_4^{-\eta_b,\eta_a}$ &
$S_4^{\eta_b,-\eta_a*}$ & $S_1^{\eta_b,-\eta_a}$ &
$S_3^{-\eta_b,-\eta_a*}$ & $S_2^{-\eta_b,-\eta_a}$ &
$S^{-\eta_b,\eta_a*}$ & $S^{\eta_b,-\eta_a}$ \\
\hline
$B_{11}$ &
$T_2^{\eta_b,\eta_a*}$ & $T_3^{\eta_b,\eta_a}$ &
$T_1^{-\eta_b,\eta_a*}$ & $T_4^{-\eta_b,\eta_a}$ &
$T_4^{\eta_b,-\eta_a*}$ & $T_1^{\eta_b,-\eta_a}$ &
$T_3^{-\eta_b,-\eta_a*}$ & $T_2^{-\eta_b,-\eta_a}$ &
$T^{-\eta_b,\eta_a*}$ & $T^{\eta_b,-\eta_a}$ \\
\hline
$C_{11}$ &
$V_2^{\eta_b,\eta_a*}$ & $V_3^{\eta_b,\eta_a}$ &
$V_1^{-\eta_b,\eta_a*}$ & $V_4^{-\eta_b,\eta_a}$ &
$V_4^{\eta_b,-\eta_a*}$ & $V_1^{\eta_b,-\eta_a}$ &
$V_3^{-\eta_b,-\eta_a*}$ & $V_2^{-\eta_b,-\eta_a}$ &
$V^{-\eta_b,\eta_a*}$ & $V^{\eta_b,-\eta_a}$ \\
\hline\hline
$D_1$ &
$1$ & $1$ & $\lambda_a\lambda_b^*$ & $1$ & $1$ &
$\lambda_a^*\lambda_b$ & $1$ & $1$ & $1$ & $1$ \\
\hline
$A_{12}$ &
$S_2^{\eta_a,\eta_b*}$ & $S_3^{\eta_a,\eta_b}$ &
$S_4^{\eta_a,-\eta_b*}$ & $S_1^{\eta_a,-\eta_b}$ &
$S_1^{-\eta_a,\eta_b*}$ & $S_4^{-\eta_a,\eta_b}$ &
$S_3^{-\eta_a,-\eta_b*}$ & $S_2^{-\eta_a,-\eta_b}$ &
$S^{-\eta_a,\eta_b*}$ & $S^{\eta_a,-\eta_b}$ \\
\hline
$B_{12}$ &
$T_2^{\eta_a,\eta_b*}$ & $T_3^{\eta_a,\eta_b}$ &
$T_4^{\eta_a,-\eta_b*}$ & $T_1^{\eta_a,-\eta_b}$ &
$T_1^{-\eta_a,\eta_b*}$ & $T_4^{-\eta_a,\eta_b}$ &
$T_3^{-\eta_a,-\eta_b*}$ & $T_2^{-\eta_a,-\eta_b}$ &
$T^{-\eta_a,\eta_b*}$ & $T^{\eta_a,-\eta_b}$ \\
\hline
$C_{12}$ &
$V_2^{\eta_a,\eta_b*}$ & $V_3^{\eta_a,\eta_b}$ &
$V_4^{\eta_a,-\eta_b*}$ & $V_1^{\eta_a,-\eta_b}$ &
$V_1^{-\eta_a,\eta_b*}$ & $V_4^{-\eta_a,\eta_b}$ &
$V_3^{-\eta_a,-\eta_b*}$ & $V_2^{-\eta_a,-\eta_b}$ &
$V^{-\eta_a,\eta_b*}$ & $V^{\eta_a,-\eta_b}$ \\
\hline\hline\hline
$N_2$ & 
$\eta_a\lambda_b$ & $\eta_a\lambda_a$ & 
$\eta_a$ & $\eta_a$ & $\eta_a$ & $\eta_a$ &
$\eta_a\lambda_a^*$ & $\eta_a\lambda_b^*$ & 
$\eta_a\lambda_b$ & $\eta_a\lambda_a$ \\
\hline\hline
$A_{21}$ &
$S_3^{\eta_b,\eta_a}$ & $S_2^{\eta_b,\eta_a*}$ &
$S_4^{-\eta_b,\eta_a}$ & $S_1^{-\eta_b,\eta_a*}$ &
$S_1^{\eta_b,-\eta_a}$ & $S_4^{\eta_b,-\eta_a*}$ &
$S_2^{-\eta_b,-\eta_a}$ & $S_3^{-\eta_b,-\eta_a*}$ &
$S^{\eta_b,-\eta_a}$ & $S^{-\eta_b,\eta_a*}$ \\
\hline
$B_{21}$ &
$T_3^{\eta_b,\eta_a}$ & $T_2^{\eta_b,\eta_a*}$ &
$T_4^{-\eta_b,\eta_a}$ & $T_1^{-\eta_b,\eta_a*}$ &
$T_1^{\eta_b,-\eta_a}$ & $T_4^{\eta_b,-\eta_a*}$ &
$T_2^{-\eta_b,-\eta_a}$ & $T_3^{-\eta_b,-\eta_a*}$ &
$T^{\eta_b,-\eta_a}$ & $T^{-\eta_b,\eta_a*}$ \\
\hline
$C_{21}$ &
$V_3^{\eta_b,\eta_a}$ & $V_2^{\eta_b,\eta_a*}$ &
$V_4^{-\eta_b,\eta_a}$ & $V_1^{-\eta_b,\eta_a*}$ &
$V_1^{\eta_b,-\eta_a}$ & $V_4^{\eta_b,-\eta_a*}$ &
$V_2^{-\eta_b,-\eta_a}$ & $V_3^{-\eta_b,-\eta_a*}$ &
$V^{\eta_b,-\eta_a}$ & $V^{-\eta_b,\eta_a*}$ \\
\hline\hline
$D_2$ &
$1$ & $1$ & $1$ & $\lambda_a\lambda_b^*$ & $\lambda_a^*\lambda_b$ &
$1$ & $1$ & $1$ & $1$ & $1$ \\
\hline
$A_{22}$ &
$S_3^{\eta_a,\eta_b}$ & $S_2^{\eta_a,\eta_b*}$ &
$S_1^{\eta_a,-\eta_b}$ & $S_4^{\eta_a,-\eta_b*}$ &
$S_4^{-\eta_a,\eta_b}$ & $S_1^{-\eta_a,\eta_b*}$ &
$S_2^{-\eta_a,-\eta_b}$ & $S_3^{-\eta_a,-\eta_b*}$ &
$S^{\eta_a,-\eta_b}$ & $S^{-\eta_a,\eta_b*}$ \\
\hline
$B_{22}$ &
$T_3^{\eta_a,\eta_b}$ & $T_2^{\eta_a,\eta_b*}$ &
$T_1^{\eta_a,-\eta_b}$ & $T_4^{\eta_a,-\eta_b*}$ &
$T_4^{-\eta_a,\eta_b}$ & $T_1^{-\eta_a,\eta_b*}$ &
$T_2^{-\eta_a,-\eta_b}$ & $T_3^{-\eta_a,-\eta_b*}$ &
$T^{\eta_a,-\eta_b}$ & $T^{-\eta_a,\eta_b*}$ \\
\hline
$C_{22}$ &
$V_3^{\eta_a,\eta_b}$ & $V_2^{\eta_a,\eta_b*}$ &
$V_1^{\eta_a,-\eta_b}$ & $V_4^{\eta_a,-\eta_b*}$ &
$V_4^{-\eta_a,\eta_b}$ & $V_1^{-\eta_a,\eta_b*}$ &
$V_2^{-\eta_a,-\eta_b}$ & $V_3^{-\eta_a,-\eta_b*}$ &
$V^{\eta_a,-\eta_b}$ & $V^{-\eta_a,\eta_b*}$ \\
\hline
\end{tabular}
\caption[]{List of the constant factors appearing in the amplitude 
(\ref{eq:ijab}) for all $(ij)(ab)$ neutrino pair production processes 
according to their labelling defined in Sect.\ref{Sect5}.}
\label{Table:ijab}
\end{center}
\end{table}

\vspace{10pt}

\begin{table}
\begin{center}
\begin{tabular}{|c||c|c|c|c|c|c|c|c||c|c|}
\hline
 & ai1 & ai2 & ai3 & ai4 & ai5 & ai6 & ai7 & ai8 & Mai1 & Mai2 \\
\hline\hline
$N_1$ & 
$\eta_a$ & $\eta_a$ & 
$\eta_a\lambda_b^*$ & $\eta_a\lambda_a^*$ & 
$\eta_a\lambda_a$ & $\eta_a\lambda_b$ &
$\eta_a$ & $\eta_a$ & 
$\eta_a$ & $\eta_a$ \\
\hline\hline
$A_{11}$ &
$S_4^{\eta_b,\eta_a*}$ & $S_1^{\eta_b,\eta_a}$ &
$S_3^{-\eta_b,\eta_a*}$ & $S_2^{-\eta_b,\eta_a}$ &
$S_2^{\eta_b,-\eta_a*}$ & $S_3^{\eta_b,-\eta_a}$ &
$S_1^{-\eta_b,-\eta_a*}$ & $S_4^{-\eta_b,-\eta_a}$ &
$S^{-\eta_b,-\eta_a*}$ & $S^{\eta_b,\eta_a}$ \\
\hline
$B_{11}$ &
$T_4^{\eta_b,\eta_a*}$ & $T_1^{\eta_b,\eta_a}$ &
$T_3^{-\eta_b,\eta_a*}$ & $T_2^{-\eta_b,\eta_a}$ &
$T_2^{\eta_b,-\eta_a*}$ & $T_3^{\eta_b,-\eta_a}$ &
$T_1^{-\eta_b,-\eta_a*}$ & $T_4^{-\eta_b,-\eta_a}$ &
$T^{-\eta_b,-\eta_a*}$ & $T^{\eta_b,\eta_a}$ \\
\hline
$C_{11}$ &
$V_4^{\eta_b,\eta_a*}$ & $V_1^{\eta_b,\eta_a}$ &
$V_3^{-\eta_b,\eta_a*}$ & $V_2^{-\eta_b,\eta_a}$ &
$V_2^{\eta_b,-\eta_a*}$ & $V_3^{\eta_b,-\eta_a}$ &
$V_1^{-\eta_b,-\eta_a*}$ & $V_4^{-\eta_b,-\eta_a}$ &
$V^{-\eta_b,-\eta_a*}$ & $V^{\eta_b,\eta_a}$ \\
\hline\hline
$D_1$ &
$1$ & $\lambda_a^*\lambda_b$ & $1$ & $1$ & $1$ &
$1$ & $\lambda_a\lambda_b^*$ & $1$ & $1$ & $\lambda_a^*\lambda_b$ \\
\hline
$A_{12}$ &
$S_1^{\eta_a,\eta_b*}$ & $S_4^{\eta_a,\eta_b}$ &
$S_3^{\eta_a,-\eta_b*}$ & $S_2^{\eta_a,-\eta_b}$ &
$S_2^{-\eta_a,\eta_b*}$ & $S_3^{-\eta_a,\eta_b}$ &
$S_4^{-\eta_a,-\eta_b*}$ & $S_1^{-\eta_a,-\eta_b}$ &
$S^{\eta_a,\eta_b*}$ & $S^{-\eta_a,-\eta_b}$ \\
\hline
$B_{12}$ &
$T_1^{\eta_a,\eta_b*}$ & $T_4^{\eta_a,\eta_b}$ &
$T_3^{\eta_a,-\eta_b*}$ & $T_2^{\eta_a,-\eta_b}$ &
$T_2^{-\eta_a,\eta_b*}$ & $T_3^{-\eta_a,\eta_b}$ &
$T_4^{-\eta_a,-\eta_b*}$ & $T_1^{-\eta_a,-\eta_b}$ &
$T^{\eta_a,\eta_b*}$ & $T^{-\eta_a,-\eta_b}$ \\
\hline
$C_{12}$ &
$V_1^{\eta_a,\eta_b*}$ & $V_4^{\eta_a,\eta_b}$ &
$V_3^{\eta_a,-\eta_b*}$ & $V_2^{\eta_a,-\eta_b}$ &
$V_2^{-\eta_a,\eta_b*}$ & $V_3^{-\eta_a,\eta_b}$ &
$V_4^{-\eta_a,-\eta_b*}$ & $V_1^{-\eta_a,-\eta_b}$ &
$V^{\eta_a,\eta_b*}$ & $V^{-\eta_a,-\eta_b}$ \\
\hline\hline\hline
$N_2$ & 
$\eta_b$ & $\eta_b$ & 
$\eta_b\lambda_a^*$ & $\eta_b\lambda_b^*$ & 
$\eta_b\lambda_b$ & $\eta_b\lambda_a$ &
$\eta_b$ & $\eta_b$ & 
$\eta_b$ & $\eta_b$ \\
\hline\hline
$A_{21}$ &
$S_1^{\eta_b,\eta_a}$ & $S_4^{\eta_b,\eta_a*}$ &
$S_2^{-\eta_b,\eta_a}$ & $S_3^{-\eta_b,\eta_a*}$ &
$S_3^{\eta_b,-\eta_a}$ & $S_2^{\eta_b,-\eta_a*}$ &
$S_4^{-\eta_b,-\eta_a}$ & $S_1^{-\eta_b,-\eta_a*}$ &
$S^{\eta_b,\eta_a}$ & $S^{-\eta_b,-\eta_a*}$ \\
\hline
$B_{21}$ &
$T_1^{\eta_b,\eta_a}$ & $T_4^{\eta_b,\eta_a*}$ &
$T_2^{-\eta_b,\eta_a}$ & $T_3^{-\eta_b,\eta_a*}$ &
$T_3^{\eta_b,-\eta_a}$ & $T_2^{\eta_b,-\eta_a*}$ &
$T_4^{-\eta_b,-\eta_a}$ & $T_1^{-\eta_b,-\eta_a*}$ &
$T^{\eta_b,\eta_a}$ & $T^{-\eta_b,-\eta_a*}$ \\
\hline
$C_{21}$ &
$V_1^{\eta_b,\eta_a}$ & $V_4^{\eta_b,\eta_a*}$ &
$V_2^{-\eta_b,\eta_a}$ & $V_3^{-\eta_b,\eta_a*}$ &
$V_3^{\eta_b,-\eta_a}$ & $V_2^{\eta_b,-\eta_a*}$ &
$V_4^{-\eta_b,-\eta_a}$ & $V_1^{-\eta_b,-\eta_a*}$ &
$V^{\eta_b,\eta_a}$ & $V^{-\eta_b,-\eta_a*}$ \\
\hline\hline
$D_2$ &
$\lambda_a^*\lambda_b$ & $1$ & $1$ & $1$ & $1$ &
$1$ & $1$ & $\lambda_a\lambda_b^*$ & $\lambda_a^*\lambda_b$ & $1$ \\
\hline
$A_{22}$ &
$S_4^{\eta_a,\eta_b}$ & $S_1^{\eta_a,\eta_b*}$ &
$S_2^{\eta_a,-\eta_b}$ & $S_3^{\eta_a,-\eta_b*}$ &
$S_3^{-\eta_a,\eta_b}$ & $S_2^{-\eta_a,\eta_b*}$ &
$S_1^{-\eta_a,-\eta_b}$ & $S_4^{-\eta_a,-\eta_b*}$ &
$S^{-\eta_a,-\eta_b}$ & $S^{\eta_a,\eta_b}$ \\
\hline
$B_{22}$ &
$T_4^{\eta_a,\eta_b}$ & $T_1^{\eta_a,\eta_b*}$ &
$T_2^{\eta_a,-\eta_b}$ & $T_3^{\eta_a,-\eta_b*}$ &
$T_3^{-\eta_a,\eta_b}$ & $T_2^{-\eta_a,\eta_b*}$ &
$T_1^{-\eta_a,-\eta_b}$ & $T_4^{-\eta_a,-\eta_b*}$ &
$T^{-\eta_a,-\eta_b}$ & $T^{\eta_a,\eta_b}$ \\
\hline
$C_{22}$ &
$V_4^{\eta_a,\eta_b}$ & $V_1^{\eta_a,\eta_b*}$ &
$V_2^{\eta_a,-\eta_b}$ & $V_3^{\eta_a,-\eta_b*}$ &
$V_3^{-\eta_a,\eta_b}$ & $V_2^{-\eta_a,\eta_b*}$ &
$V_1^{-\eta_a,-\eta_b}$ & $V_4^{-\eta_a,-\eta_b*}$ &
$V^{-\eta_a,-\eta_b}$ & $V^{\eta_a,\eta_b}$ \\
\hline
\end{tabular}
\caption[]{List of the constant factors appearing in the amplitude 
(\ref{eq:aibj}) for all $(ai)(bj)$ neutrino scattering processes according 
to their labelling defined in Sect.\ref{Subsect6.1}.}
\label{Table:aibj}
\end{center}
\end{table}

\vspace{10pt}

\begin{table}
\begin{center}
\begin{tabular}{|c||c|c|c|c|c|c|c|c||c|c|}
\hline
 & aj1 & aj2 & aj3 & aj4 & aj5 & aj6 & aj7 & aj8 & Maj1 & Maj2 \\
\hline\hline
$N_1$ & 
$\eta_b$ & $\eta_b$ & 
$\eta_b\lambda_a^*$ & $\eta_b\lambda_b^*$ & 
$\eta_b\lambda_b$ & $\eta_b\lambda_a$ &
$\eta_b$ & $\eta_b$ & 
$\eta_b$ & $\eta_b$ \\
\hline\hline
$A_{11}$ &
$S_1^{\eta_b,\eta_a}$ & $S_4^{\eta_b,\eta_a*}$ &
$S_2^{-\eta_b,\eta_a}$ & $S_3^{-\eta_b,\eta_a*}$ &
$S_3^{\eta_b,-\eta_a}$ & $S_2^{\eta_b,-\eta_a*}$ &
$S_4^{-\eta_b,-\eta_a}$ & $S_1^{-\eta_b,-\eta_a*}$ &
$S^{\eta_b,\eta_a}$ & $S^{-\eta_b,-\eta_a*}$ \\
\hline
$B_{11}$ &
$T_1^{\eta_b,\eta_a}$ & $T_4^{\eta_b,\eta_a*}$ &
$T_2^{-\eta_b,\eta_a}$ & $T_3^{-\eta_b,\eta_a*}$ &
$T_3^{\eta_b,-\eta_a}$ & $T_2^{\eta_b,-\eta_a*}$ &
$T_4^{-\eta_b,-\eta_a}$ & $T_1^{-\eta_b,-\eta_a*}$ &
$T^{\eta_b,\eta_a}$ & $T^{-\eta_b,-\eta_a*}$ \\
\hline
$C_{11}$ &
$V_1^{\eta_b,\eta_a}$ & $V_4^{\eta_b,\eta_a*}$ &
$V_2^{-\eta_b,\eta_a}$ & $V_3^{-\eta_b,\eta_a*}$ &
$V_3^{\eta_b,-\eta_a}$ & $V_2^{\eta_b,-\eta_a*}$ &
$V_4^{-\eta_b,-\eta_a}$ & $V_1^{-\eta_b,-\eta_a*}$ &
$V^{\eta_b,\eta_a}$ & $V^{-\eta_b,-\eta_a*}$ \\
\hline\hline
$D_1$ &
$\lambda_a^*\lambda_b$ & $1$ & $1$ & $1$ & $1$ &
$1$ & $1$ & $\lambda_a\lambda_b^*$ & $\lambda_a^*\lambda_b$ & $1$ \\
\hline
$A_{12}$ &
$S_4^{\eta_a,\eta_b}$ & $S_1^{\eta_a,\eta_b*}$ &
$S_2^{\eta_a,-\eta_b}$ & $S_3^{\eta_a,-\eta_b*}$ &
$S_3^{-\eta_a,\eta_b}$ & $S_2^{-\eta_a,\eta_b*}$ &
$S_1^{-\eta_a,-\eta_b}$ & $S_4^{-\eta_a,-\eta_b*}$ &
$S^{-\eta_a,-\eta_b}$ & $S^{\eta_a,\eta_b*}$ \\
\hline
$B_{12}$ &
$T_4^{\eta_a,\eta_b}$ & $T_1^{\eta_a,\eta_b*}$ &
$T_2^{\eta_a,-\eta_b}$ & $T_3^{\eta_a,-\eta_b*}$ &
$T_3^{-\eta_a,\eta_b}$ & $T_2^{-\eta_a,\eta_b*}$ &
$T_1^{-\eta_a,-\eta_b}$ & $T_4^{-\eta_a,-\eta_b*}$ &
$T^{-\eta_a,-\eta_b}$ & $T^{\eta_a,\eta_b*}$ \\
\hline
$C_{12}$ &
$V_4^{\eta_a,\eta_b}$ & $V_1^{\eta_a,\eta_b*}$ &
$V_2^{\eta_a,-\eta_b}$ & $V_3^{\eta_a,-\eta_b*}$ &
$V_3^{-\eta_a,\eta_b}$ & $V_2^{-\eta_a,\eta_b*}$ &
$V_1^{-\eta_a,-\eta_b}$ & $V_4^{-\eta_a,-\eta_b*}$ &
$V^{-\eta_a,-\eta_b}$ & $V^{\eta_a,\eta_b*}$ \\
\hline\hline\hline
$N_2$ & 
$\eta_a$ & $\eta_a$ & 
$\eta_a\lambda_b^*$ & $\eta_a\lambda_a^*$ & 
$\eta_a\lambda_a$ & $\eta_a\lambda_b$ &
$\eta_a$ & $\eta_a$ & 
$\eta_a$ & $\eta_a$ \\
\hline\hline
$A_{21}$ &
$S_4^{\eta_b,\eta_a*}$ & $S_1^{\eta_b,\eta_a}$ &
$S_3^{-\eta_b,\eta_a*}$ & $S_2^{-\eta_b,\eta_a}$ &
$S_2^{\eta_b,-\eta_a*}$ & $S_3^{\eta_b,-\eta_a}$ &
$S_1^{-\eta_b,-\eta_a*}$ & $S_4^{-\eta_b,-\eta_a}$ &
$S^{-\eta_b,-\eta_a*}$ & $S^{\eta_b,\eta_a}$ \\
\hline
$B_{21}$ &
$T_4^{\eta_b,\eta_a*}$ & $T_1^{\eta_b,\eta_a}$ &
$T_3^{-\eta_b,\eta_a*}$ & $T_2^{-\eta_b,\eta_a}$ &
$T_2^{\eta_b,-\eta_a*}$ & $T_3^{\eta_b,-\eta_a}$ &
$T_1^{-\eta_b,-\eta_a*}$ & $T_4^{-\eta_b,-\eta_a}$ &
$T^{-\eta_b,-\eta_a*}$ & $T^{\eta_b,\eta_a}$ \\
\hline
$C_{21}$ &
$V_4^{\eta_b,\eta_a*}$ & $V_1^{\eta_b,\eta_a}$ &
$V_3^{-\eta_b,\eta_a*}$ & $V_2^{-\eta_b,\eta_a}$ &
$V_2^{\eta_b,-\eta_a*}$ & $V_3^{\eta_b,-\eta_a}$ &
$V_1^{-\eta_b,-\eta_a*}$ & $V_4^{-\eta_b,-\eta_a}$ &
$V^{-\eta_b,-\eta_a*}$ & $V^{\eta_b,\eta_a}$ \\
\hline\hline
$D_2$ &
$1$ & $\lambda_a^*\lambda_b$ & $1$ & $1$ & $1$ &
$1$ & $\lambda_a\lambda_b^*$ & $1$ & $1$ & $\lambda_a^*\lambda_b$ \\
\hline
$A_{22}$ &
$S_1^{\eta_a,\eta_b*}$ & $S_4^{\eta_a,\eta_b}$ &
$S_3^{\eta_a,-\eta_b*}$ & $S_2^{\eta_a,-\eta_b}$ &
$S_2^{-\eta_a,\eta_b*}$ & $S_3^{-\eta_a,\eta_b}$ &
$S_4^{-\eta_a,-\eta_b*}$ & $S_1^{-\eta_a,-\eta_b}$ &
$S^{\eta_a,\eta_b*}$ & $S^{-\eta_a,-\eta_b}$ \\
\hline
$B_{22}$ &
$T_1^{\eta_a,\eta_b*}$ & $T_4^{\eta_a,\eta_b}$ &
$T_3^{\eta_a,-\eta_b*}$ & $T_2^{\eta_a,-\eta_b}$ &
$T_2^{-\eta_a,\eta_b*}$ & $T_3^{-\eta_a,\eta_b}$ &
$T_4^{-\eta_a,-\eta_b*}$ & $T_1^{-\eta_a,-\eta_b}$ &
$T^{\eta_a,\eta_b*}$ & $T^{-\eta_a,-\eta_b}$ \\
\hline
$C_{22}$ &
$V_1^{\eta_a,\eta_b*}$ & $V_4^{\eta_a,\eta_b}$ &
$V_3^{\eta_a,-\eta_b*}$ & $V_2^{\eta_a,-\eta_b}$ &
$V_2^{-\eta_a,\eta_b*}$ & $V_3^{-\eta_a,\eta_b}$ &
$V_4^{-\eta_a,-\eta_b*}$ & $V_1^{-\eta_a,-\eta_b}$ &
$V^{\eta_a,\eta_b*}$ & $V^{-\eta_a,-\eta_b}$ \\
\hline
\end{tabular}
\caption[]{List of the constant factors appearing in the amplitude 
(\ref{eq:ajbi}) for all $(aj)(bi)$ neutrino scattering processes according 
to their labelling defined in Sect.\ref{Subsect6.2}.}
\label{Table:ajbi}
\end{center}
\end{table}

\vspace{10pt}

\begin{table}
\begin{center}
\begin{tabular}{|c||c|c|c|c|c|c|c|c||c|c|}
\hline
 & bi1 & bi2 & bi3 & bi4 & bi5 & bi6 & bi7 & bi8 & Mbi1 & Mbi2 \\
\hline\hline
$N_1$ & 
$\eta_a$ & $\eta_a$ & 
$\eta_a\lambda_a$ & $\eta_a\lambda_b$ & 
$\eta_a\lambda_b^*$ & $\eta_a\lambda_a^*$ &
$\eta_a$ & $\eta_a$ & 
$\eta_a$ & $\eta_a$ \\
\hline\hline
$A_{11}$ &
$S_1^{\eta_b,\eta_a*}$ & $S_4^{\eta_b,\eta_a}$ &
$S_2^{-\eta_b,\eta_a*}$ & $S_3^{-\eta_b,\eta_a}$ &
$S_3^{\eta_b,-\eta_a*}$ & $S_2^{\eta_b,-\eta_a}$ &
$S_4^{-\eta_b,-\eta_a*}$ & $S_1^{-\eta_b,-\eta_a}$ &
$S^{\eta_b,\eta_a*}$ & $S^{-\eta_b,-\eta_a}$ \\
\hline
$B_{11}$ &
$T_1^{\eta_b,\eta_a*}$ & $T_4^{\eta_b,\eta_a}$ &
$T_2^{-\eta_b,\eta_a*}$ & $T_3^{-\eta_b,\eta_a}$ &
$T_3^{\eta_b,-\eta_a*}$ & $T_2^{\eta_b,-\eta_a}$ &
$T_4^{-\eta_b,-\eta_a*}$ & $T_1^{-\eta_b,-\eta_a}$ &
$T^{\eta_b,\eta_a*}$ & $T^{-\eta_b,-\eta_a}$ \\
\hline
$C_{11}$ &
$V_1^{\eta_b,\eta_a*}$ & $V_4^{\eta_b,\eta_a}$ &
$V_2^{-\eta_b,\eta_a*}$ & $V_3^{-\eta_b,\eta_a}$ &
$V_3^{\eta_b,-\eta_a*}$ & $V_2^{\eta_b,-\eta_a}$ &
$V_4^{-\eta_b,-\eta_a*}$ & $V_1^{-\eta_b,-\eta_a}$ &
$V^{\eta_b,\eta_a*}$ & $V^{-\eta_b,-\eta_a}$ \\
\hline\hline
$D_1$ &
$\lambda_a\lambda_b^*$ & $1$ & $1$ & $1$ & $1$ &
$1$ & $1$ & $\lambda_a^*\lambda_b$ & $\lambda_a\lambda_b^*$ & $1$ \\
\hline
$A_{12}$ &
$S_4^{\eta_a,\eta_b*}$ & $S_1^{\eta_a,\eta_b}$ &
$S_2^{\eta_a,-\eta_b*}$ & $S_3^{\eta_a,-\eta_b}$ &
$S_3^{-\eta_a,\eta_b*}$ & $S_2^{-\eta_a,\eta_b}$ &
$S_1^{-\eta_a,-\eta_b*}$ & $S_4^{-\eta_a,-\eta_b}$ &
$S^{-\eta_a,-\eta_b*}$ & $S^{\eta_a,\eta_b}$ \\
\hline
$B_{12}$ &
$T_4^{\eta_a,\eta_b*}$ & $T_1^{\eta_a,\eta_b}$ &
$T_2^{\eta_a,-\eta_b*}$ & $T_3^{\eta_a,-\eta_b}$ &
$T_3^{-\eta_a,\eta_b*}$ & $T_2^{-\eta_a,\eta_b}$ &
$T_1^{-\eta_a,-\eta_b*}$ & $T_4^{-\eta_a,-\eta_b}$ &
$T^{-\eta_a,-\eta_b*}$ & $T^{\eta_a,\eta_b}$ \\
\hline
$C_{12}$ &
$V_4^{\eta_a,\eta_b*}$ & $V_1^{\eta_a,\eta_b}$ &
$V_2^{\eta_a,-\eta_b*}$ & $V_3^{\eta_a,-\eta_b}$ &
$V_3^{-\eta_a,\eta_b*}$ & $V_2^{-\eta_a,\eta_b}$ &
$V_1^{-\eta_a,-\eta_b*}$ & $V_4^{-\eta_a,-\eta_b}$ &
$V^{-\eta_a,-\eta_b*}$ & $V^{\eta_a,\eta_b}$ \\
\hline\hline\hline
$N_2$ & 
$\eta_b$ & $\eta_b$ & 
$\eta_b\lambda_b$ & $\eta_b\lambda_a$ & 
$\eta_b\lambda_a^*$ & $\eta_b\lambda_b^*$ &
$\eta_b$ & $\eta_b$ & 
$\eta_b$ & $\eta_b$ \\
\hline\hline
$A_{21}$ &
$S_4^{\eta_b,\eta_a}$ & $S_1^{\eta_b,\eta_a*}$ &
$S_3^{-\eta_b,\eta_a}$ & $S_2^{-\eta_b,\eta_a*}$ &
$S_2^{\eta_b,-\eta_a}$ & $S_3^{\eta_b,-\eta_a*}$ &
$S_1^{-\eta_b,-\eta_a}$ & $S_4^{-\eta_b,-\eta_a*}$ &
$S^{-\eta_b,-\eta_a}$ & $S^{\eta_b,\eta_a*}$ \\
\hline
$B_{21}$ &
$T_4^{\eta_b,\eta_a}$ & $T_1^{\eta_b,\eta_a*}$ &
$T_3^{-\eta_b,\eta_a}$ & $T_2^{-\eta_b,\eta_a*}$ &
$T_2^{\eta_b,-\eta_a}$ & $T_3^{\eta_b,-\eta_a*}$ &
$T_1^{-\eta_b,-\eta_a}$ & $T_4^{-\eta_b,-\eta_a*}$ &
$T^{-\eta_b,-\eta_a}$ & $T^{\eta_b,\eta_a*}$ \\
\hline
$C_{21}$ &
$V_4^{\eta_b,\eta_a}$ & $V_1^{\eta_b,\eta_a*}$ &
$V_3^{-\eta_b,\eta_a}$ & $V_2^{-\eta_b,\eta_a*}$ &
$V_2^{\eta_b,-\eta_a}$ & $V_3^{\eta_b,-\eta_a*}$ &
$V_1^{-\eta_b,-\eta_a}$ & $V_4^{-\eta_b,-\eta_a*}$ &
$V^{-\eta_b,-\eta_a}$ & $V^{\eta_b,\eta_a*}$ \\
\hline\hline
$D_2$ &
$1$ & $\lambda_a\lambda_b^*$ & $1$ & $1$ & $1$ &
$1$ & $\lambda_a^*\lambda_b$ & $1$ & $1$ & $\lambda_a\lambda_b^*$ \\
\hline
$A_{22}$ &
$S_1^{\eta_a,\eta_b}$ & $S_4^{\eta_a,\eta_b*}$ &
$S_3^{\eta_a,-\eta_b}$ & $S_2^{\eta_a,-\eta_b*}$ &
$S_2^{-\eta_a,\eta_b}$ & $S_3^{-\eta_a,\eta_b*}$ &
$S_4^{-\eta_a,-\eta_b}$ & $S_1^{-\eta_a,-\eta_b*}$ &
$S^{\eta_a,\eta_b}$ & $S^{-\eta_a,-\eta_b*}$ \\
\hline
$B_{22}$ &
$T_1^{\eta_a,\eta_b}$ & $T_4^{\eta_a,\eta_b*}$ &
$T_3^{\eta_a,-\eta_b}$ & $T_2^{\eta_a,-\eta_b*}$ &
$T_2^{-\eta_a,\eta_b}$ & $T_3^{-\eta_a,\eta_b*}$ &
$T_4^{-\eta_a,-\eta_b}$ & $T_1^{-\eta_a,-\eta_b*}$ &
$T^{\eta_a,\eta_b}$ & $T^{-\eta_a,-\eta_b*}$ \\
\hline
$C_{22}$ &
$V_1^{\eta_a,\eta_b}$ & $V_4^{\eta_a,\eta_b*}$ &
$V_3^{\eta_a,-\eta_b}$ & $V_2^{\eta_a,-\eta_b*}$ &
$V_2^{-\eta_a,\eta_b}$ & $V_3^{-\eta_a,\eta_b*}$ &
$V_4^{-\eta_a,-\eta_b}$ & $V_1^{-\eta_a,-\eta_b*}$ &
$V^{\eta_a,\eta_b}$ & $V^{-\eta_a,-\eta_b*}$ \\
\hline
\end{tabular}
\caption[]{List of the constant factors appearing in the amplitude 
(\ref{eq:biaj}) for all $(bi)(aj)$ neutrino scattering processes according 
to their labelling defined in Sect.\ref{Subsect6.3}.}
\label{Table:biaj}
\end{center}
\end{table}

\vspace{10pt}

\begin{table}
\begin{center}
\begin{tabular}{|c||c|c|c|c|c|c|c|c||c|c|}
\hline
 & bj1 & bj2 & bj3 & bj4 & bj5 & bj6 & bj7 & bj8 & Mbj1 & Mbj2 \\
\hline\hline
$N_1$ & 
$\eta_b$ & $\eta_b$ & 
$\eta_b\lambda_b$ & $\eta_b\lambda_a$ & 
$\eta_b\lambda_a^*$ & $\eta_b\lambda_b^*$ &
$\eta_b$ & $\eta_b$ & 
$\eta_b$ & $\eta_b$ \\
\hline\hline
$A_{11}$ &
$S_4^{\eta_b,\eta_a}$ & $S_1^{\eta_b,\eta_a*}$ &
$S_3^{-\eta_b,\eta_a}$ & $S_2^{-\eta_b,\eta_a*}$ &
$S_2^{\eta_b,-\eta_a}$ & $S_3^{\eta_b,-\eta_a*}$ &
$S_1^{-\eta_b,-\eta_a}$ & $S_4^{-\eta_b,-\eta_a*}$ &
$S^{-\eta_b,-\eta_a}$ & $S^{\eta_b,\eta_a*}$ \\
\hline
$B_{11}$ &
$T_4^{\eta_b,\eta_a}$ & $T_1^{\eta_b,\eta_a*}$ &
$T_3^{-\eta_b,\eta_a}$ & $T_2^{-\eta_b,\eta_a*}$ &
$T_2^{\eta_b,-\eta_a}$ & $T_3^{\eta_b,-\eta_a*}$ &
$T_1^{-\eta_b,-\eta_a}$ & $T_4^{-\eta_b,-\eta_a*}$ &
$T^{-\eta_b,-\eta_a}$ & $T^{\eta_b,\eta_a*}$ \\
\hline
$C_{11}$ &
$V_4^{\eta_b,\eta_a}$ & $V_1^{\eta_b,\eta_a*}$ &
$V_3^{-\eta_b,\eta_a}$ & $V_2^{-\eta_b,\eta_a*}$ &
$V_2^{\eta_b,-\eta_a}$ & $V_3^{\eta_b,-\eta_a*}$ &
$V_1^{-\eta_b,-\eta_a}$ & $V_4^{-\eta_b,-\eta_a*}$ &
$V^{-\eta_b,-\eta_a}$ & $V^{\eta_b,\eta_a*}$ \\
\hline\hline
$D_1$ &
$1$ & $\lambda_a\lambda_b^*$ & $1$ & $1$ & $1$ &
$1$ & $\lambda_a^*\lambda_b$ & $1$ & $1$ & $\lambda_a\lambda_b^*$ \\
\hline
$A_{12}$ &
$S_1^{\eta_a,\eta_b}$ & $S_4^{\eta_a,\eta_b*}$ &
$S_3^{\eta_a,-\eta_b}$ & $S_2^{\eta_a,-\eta_b*}$ &
$S_2^{-\eta_a,\eta_b}$ & $S_3^{-\eta_a,\eta_b*}$ &
$S_4^{-\eta_a,-\eta_b}$ & $S_1^{-\eta_a,-\eta_b*}$ &
$S^{\eta_a,\eta_b}$ & $S^{-\eta_a,-\eta_b*}$ \\
\hline
$B_{12}$ &
$T_1^{\eta_a,\eta_b}$ & $T_4^{\eta_a,\eta_b*}$ &
$T_3^{\eta_a,-\eta_b}$ & $T_2^{\eta_a,-\eta_b*}$ &
$T_2^{-\eta_a,\eta_b}$ & $T_3^{-\eta_a,\eta_b*}$ &
$T_4^{-\eta_a,-\eta_b}$ & $T_1^{-\eta_a,-\eta_b*}$ &
$T^{\eta_a,\eta_b}$ & $T^{-\eta_a,-\eta_b*}$ \\
\hline
$C_{12}$ &
$V_1^{\eta_a,\eta_b}$ & $V_4^{\eta_a,\eta_b*}$ &
$V_3^{\eta_a,-\eta_b}$ & $V_2^{\eta_a,-\eta_b*}$ &
$V_2^{-\eta_a,\eta_b}$ & $V_3^{-\eta_a,\eta_b*}$ &
$V_4^{-\eta_a,-\eta_b}$ & $V_1^{-\eta_a,-\eta_b*}$ &
$V^{\eta_a,\eta_b}$ & $V^{-\eta_a,-\eta_b*}$ \\
\hline\hline\hline
$N_2$ & 
$\eta_a$ & $\eta_a$ & 
$\eta_a\lambda_a$ & $\eta_a\lambda_b$ & 
$\eta_a\lambda_b^*$ & $\eta_a\lambda_a^*$ &
$\eta_a$ & $\eta_a$ & 
$\eta_a$ & $\eta_a$ \\
\hline\hline
$A_{21}$ &
$S_1^{\eta_b,\eta_a*}$ & $S_4^{\eta_b,\eta_a}$ &
$S_2^{-\eta_b,\eta_a*}$ & $S_3^{-\eta_b,\eta_a}$ &
$S_3^{\eta_b,-\eta_a*}$ & $S_2^{\eta_b,-\eta_a}$ &
$S_4^{-\eta_b,-\eta_a*}$ & $S_1^{-\eta_b,-\eta_a}$ &
$S^{\eta_b,\eta_a*}$ & $S^{-\eta_b,-\eta_a}$ \\
\hline
$B_{21}$ &
$T_1^{\eta_b,\eta_a*}$ & $T_4^{\eta_b,\eta_a}$ &
$T_2^{-\eta_b,\eta_a*}$ & $T_3^{-\eta_b,\eta_a}$ &
$T_3^{\eta_b,-\eta_a*}$ & $T_2^{\eta_b,-\eta_a}$ &
$T_4^{-\eta_b,-\eta_a*}$ & $T_1^{-\eta_b,-\eta_a}$ &
$T^{\eta_b,\eta_a*}$ & $T^{-\eta_b,-\eta_a}$ \\
\hline
$C_{21}$ &
$V_1^{\eta_b,\eta_a*}$ & $V_4^{\eta_b,\eta_a}$ &
$V_2^{-\eta_b,\eta_a*}$ & $V_3^{-\eta_b,\eta_a}$ &
$V_3^{\eta_b,-\eta_a*}$ & $V_2^{\eta_b,-\eta_a}$ &
$V_4^{-\eta_b,-\eta_a*}$ & $V_1^{-\eta_b,-\eta_a}$ &
$V^{\eta_b,\eta_a*}$ & $V^{-\eta_b,-\eta_a}$ \\
\hline\hline
$D_2$ &
$\lambda_a\lambda_b^*$ & $1$ & $1$ & $1$ & $1$ &
$1$ & $1$ & $\lambda_a^*\lambda_b$ & $\lambda_a\lambda_b^*$ & $1$ \\
\hline
$A_{22}$ &
$S_4^{\eta_a,\eta_b*}$ & $S_1^{\eta_a,\eta_b}$ &
$S_2^{\eta_a,-\eta_b*}$ & $S_3^{\eta_a,-\eta_b}$ &
$S_3^{-\eta_a,\eta_b*}$ & $S_2^{-\eta_a,\eta_b}$ &
$S_1^{-\eta_a,-\eta_b*}$ & $S_4^{-\eta_a,-\eta_b}$ &
$S^{-\eta_a,-\eta_b*}$ & $S^{\eta_a,\eta_b}$ \\
\hline
$B_{22}$ &
$T_4^{\eta_a,\eta_b*}$ & $T_1^{\eta_a,\eta_b}$ &
$T_2^{\eta_a,-\eta_b*}$ & $T_3^{\eta_a,-\eta_b}$ &
$T_3^{-\eta_a,\eta_b*}$ & $T_2^{-\eta_a,\eta_b}$ &
$T_1^{-\eta_a,-\eta_b*}$ & $T_4^{-\eta_a,-\eta_b}$ &
$T^{-\eta_a,-\eta_b*}$ & $T^{\eta_a,\eta_b}$ \\
\hline
$C_{22}$ &
$V_4^{\eta_a,\eta_b*}$ & $V_1^{\eta_a,\eta_b}$ &
$V_2^{\eta_a,-\eta_b*}$ & $V_3^{\eta_a,-\eta_b}$ &
$V_3^{-\eta_a,\eta_b*}$ & $V_2^{-\eta_a,\eta_b}$ &
$V_1^{-\eta_a,-\eta_b*}$ & $V_4^{-\eta_a,-\eta_b}$ &
$V^{-\eta_a,-\eta_b*}$ & $V^{\eta_a,\eta_b}$ \\
\hline
\end{tabular}
\caption[]{List of the constant factors appearing in the amplitude 
(\ref{eq:bjai}) for all $(bj)(ai)$ neutrino scattering processes according 
to their labelling defined in Sect.\ref{Subsect6.4}.}
\label{Table:bjai}
\end{center}
\end{table}

\end{document}